\documentclass{article}

    \usepackage[final]{Styles/neurips_2024}

\usepackage[utf8]{inputenc} 
\usepackage[T1]{fontenc}    
\usepackage{hyperref}       
\usepackage{url}            
\usepackage{booktabs}       
\usepackage{amsfonts}       
\usepackage{nicefrac}       
\usepackage{microtype}      
\usepackage{xcolor}         
\usepackage{subfiles} 
\usepackage{microtype}
\usepackage{graphicx}
\usepackage{subfigure}
\usepackage{booktabs, siunitx} 
\usepackage{pdfpages}
\usepackage{hyperref}
\usepackage{array}
\usepackage{placeins}
\usepackage{stfloats}
\usepackage{amsmath}
\usepackage{amssymb}
\usepackage{mathtools}
\usepackage{amsthm}
\usepackage{thmtools}
\usepackage{thm-restate}
\usepackage{subcaption}
\usepackage{hyperref}
\usepackage{graphicx}
\usepackage{rotating} 
\usepackage[capitalize, noabbrev]{cleveref}

\usepackage{booktabs}
\usepackage{multirow}
\usepackage{array}
\usepackage{amsmath}  

\usepackage{multirow}
\usepackage[utf8]{inputenc} 
\usepackage[T1]{fontenc}    
\usepackage{hyperref}       
\usepackage{url}            
\usepackage{booktabs}       
\usepackage{amsfonts}       
\usepackage{nicefrac}       
\usepackage{microtype}      
\usepackage{xcolor}         

\title{FINALLY: fast and universal speech enhancement with studio-like quality}

\author{
    Nicholas Babaev$^*$  \\ Samsung Research
    \And 
    Kirill Tamogashev$^*$ \\ Samsung Research
    \And 
    Azat Saginbaev  \\ Samsung Research
    \And 
    Ivan  Shchekotov \\ Samsung Research
    \AND 
    Hanbin Bae \\ Samsung Research
    \And 
    Hosang Sung \\ Samsung Research
    \And 
    WonJun Lee \\ Samsung Reseach 
    \And 
    Hoon-Young Cho \\ Samsung Research
    \And 
    Pavel Andreev\thanks{
    Equal contribution. Correspondence to p.andreev@samsung.com.
    } \\ Samsung Research
}

\begin{document}

\maketitle

\begin{abstract}
In this paper, we address the challenge of speech enhancement in real-world recordings, which often contain various forms of distortion, such as background noise, reverberation, and microphone artefacts.
We revisit the use of Generative Adversarial Networks (GANs) for speech enhancement and theoretically show that GANs are naturally inclined to seek the point of maximum density within the conditional clean speech distribution, which, as we argue, is essential for the speech enhancement task.
We study various feature extractors for perceptual loss to facilitate the stability of adversarial training, developing a methodology for probing the structure of the feature space.
This leads us to integrate WavLM-based perceptual loss into MS-STFT adversarial training pipeline, creating an effective and stable training procedure for the speech enhancement model.
The resulting speech enhancement model, which we refer to as FINALLY, builds upon the HiFi++ architecture, augmented with a WavLM encoder and a novel training pipeline.
Empirical results on various datasets confirm our model's ability to produce clear, high-quality speech at 48 kHz, achieving state-of-the-art performance in the field of speech enhancement. Demo page: \href{https://samsunglabs.github.io/FINALLY-page/}{https://samsunglabs.github.io/FINALLY-page/}
\end{abstract}

\section{Introduction}

Speech recordings are often contaminated with background noise, reverberation, reduced frequency bandwidth, and other distortions. Unlike classical speech enhancement \citep{ephraim1984speech, pascual2017segan}, which considers each task separately, universal speech enhancement \citep{serra2022universal, su2021hifi, liu2022voicefixer} aims to restore speech from all types of distortions simultaneously. Thus, universal speech enhancement seeks to generalize across a wide range of distortions, making it more suitable for real-world applications where multiple distortions may coexist.

Recent studies have categorized the problem of speech enhancement as a task of learning the clean speech distribution conditioned on degraded signals \citep{lemercier2023storm, serra2022universal, richter2023speech}. This problem is often addressed using diffusion models \citep{ho2020denoising, song2020score}, which are renowned for their exceptional ability to learn distributions. Diffusion models have recently achieved state-of-the-art results in universal speech enhancement \citep{serra2022universal}. However, the impressive performance of diffusion models comes with the high computational cost of their iterative inference process.

It is important to note that the speech enhancement problem does not require the model to learn the entire conditional distribution. In practice, when presented with a noisy speech sample, the goal is often to obtain the most probable clean speech sample that retains the lexical content and voice of the original. This contrasts with applications such as text-to-image synthesis \citep{ramesh2022hierarchical, lee2024holistic, rombach2022high}, where the objective is to generate a variety of images for each text prompt due to the higher level of uncertainty and the need for diverse options to select the best image. For most speech enhancement applications, such as voice calls and compensation for poor recording conditions, capturing the entire conditional distribution is not necessary. Instead, it is more important to retrieve the most likely sample of this distribution (the main mode), which might be a simpler task.

Diffusion models' main advantage over generative adversarial networks (GANs) \citep{goodfellow2014generative} is their ability to capture different modes of the distribution. However, we argue that this property is not typically required for the task of speech enhancement and may unnecessarily complicate the operation of the neural network. Conversely, we show that GANs tend to retrieve the main mode of the distribution—precisely what speech enhancement should typically do.

Therefore, in this work, we revisit the GAN framework for speech enhancement and demonstrate that it provides rapid and high-quality universal speech enhancement. Our model outperforms both diffusion models and previous GAN-based models, achieving an unprecedented level of quality on both simulated and real-world data.

Our main contributions are as follows:

\begin{enumerate}
    \item We theoretically analyse the adversarial training with the least squares GAN (LS-GAN) loss and demonstrate that a generator predicting a single sample per input condition (producing a conditional distribution that is a delta function) is incentivized to select the point of maximum density. Therefore, we establish that LS-GAN training can implicitly regress for the main mode of the distribution, aligning with the objectives of the speech enhancement problem.
    \item We investigate various feature extractors as backbones for perceptual loss and propose criteria for selecting an extractor based on the structure of its feature space. These criteria are validated by empirical results from a neural vocoding task, indicating that the convolutional features of the WavLM neural network\citep{chen2022wavlm} are well-suited for perceptual loss in speech generation.
    \item We develop a novel model for universal speech enhancement that integrates the proposed perceptual loss with MS-STFT discriminator training~\citep{defossez2023high} and enhances the architecture of the HiFi++ generator \citep{andreev2022hifi++} by combining it with a self-supervised pre-trained WavLM encoder \citep{chen2022wavlm}. Our final model delivers state-of-the-art performance on real-world data, producing high-quality, studio-like speech at 48 kHz.
\end{enumerate}
\label{method}

\newcommand{\mlcell}[2][p{2cm}c]{%
    \begin{tabular}[#1]{@{}c@{}}#2\end{tabular}
}
\newcolumntype{H}{>{\setbox0=\hbox\bgroup}c<{\egroup}@{}}

\section{Mode Collapse and Speech Enhancement}

The first question that we address is \textbf{what is the practical purpose of a speech enhancement model}. The practical goal of a speech enhancement model is to restore the audio signal containing the speech characteristics of the original recording, including the voice, linguistic content, and prosody. Thus, loosely speaking, the purpose of the speech enhancement task for many applications is not “generative” in its essence, in the sense that the speech enhancement model should not generate new speech content but rather “refine” existing speech as if it was recorded in ideal conditions (studio-like quality). From the mathematical point of view, this means that the speech enhancement model should retrieve the most probable reconstruction of the clean speech \( y \) given the corrupted version \( x \), i.e., \( y = \mathrm{arg\,max}_y \ p_{\text{clean}}(y|x) \).

This formulation re-considers the probabilistic speech enhancement formulation, which is widely used in the literature. In such formulation, the speech enhancement model is aimed to capture the entire conditional distribution \( p_{\text{clean}}(y|x) \). This formulation might be especially appealing in situations with high generation ambiguity, e.g., a low SNR scenario where clean speech content could not be restored unambiguously. In this case, the speech enhancement model could be used to generate multiple reconstructions, the best of which is then selected by the end user. However, we note that this formulation might be redundant and not particularly relevant for many practical applications since the ambiguity in generation can be resolved by more straightforward means such as conditioning on linguistic content \citep{koizumi2023miipher}.

In practice, for many applications, a more natural way of formalizing speech enhancement is to treat it as a regression problem which aims at predicting the point of highest probability of the conditional distribution \( \mathrm{arg\,max}_y \ p_{\text{clean}}(y|x) \). This formulation has the advantage of simplifying the task, since finding the highest mode of the distribution might be significantly easier than learning the entire distribution. Therefore, the speech enhancement models built for this formulation are likely to be more efficient after deployment since they solve a simpler task. We note that in the context of speech enhancement, the speed of inference is always of major concern in practice.

Given this formulation, we argue that the framework of generative adversarial networks (GANs) is more naturally suited for the speech enhancement problem than diffusion models. We show that GAN training naturally leads to the mode-seeking behaviour of the generator, which aligns with the formulation introduced above. Additionally, GANs enjoy one forward pass inference, which is in contrast to the iterative nature of diffusion models.

Let \( p_g(y|x) \) be a family of waveform distributions produced by the generator \( g_\theta(x) \). \citet{mao2017least} showed that training with Least Squares GAN (LS-GAN) leads to the minimization of the Pearson \(\chi^2\) divergence \( \chi^2_{\text{Pearson}} \left( p_g \middle\| \frac{p_{\text{clean}} + p_g}{2} \right) \). We propose that if \( p_g(y|x) \) approaches \( \delta(y - g_\theta(x)) \) under some parametrization, the minimization of this divergence leads to \( g_\theta(x) = \mathrm{arg\,max}_y\ p_{\text{clean}}(y|x) \). This means that if the generator deterministically predicts the clean waveform from the degraded signal, the LS-GAN loss encourages the generator to predict the point of maximum \( p_{\text{clean}}(y|x) \) density. We note that although prior work by \citep{li2023mode} demonstrated the mode-covering property for the optimization of Pearson \(\chi^2\) divergence, our result pertains to a deterministic generator setting, which is outside the scope of analysis provided by \citet{li2023mode}.

To prove this result, we consider the delta function as a limit of indicator density functions \( p^{\xi}_g(y|x) = \xi^n / 2^n \cdot \mathbf{1}_{y - g_\theta(x) \in [- 1/\xi, 1/\xi]^n} \), i.e., \( p^{\xi}_g(y|x) = \xi^n/2^n \) if \( y - g_\theta(x) \in [ - 1/\xi, 1/\xi]^n \) and 0 otherwise, where \( n \) is the number of dimensions of \( y \). Note that \( \int p^{\xi}_g(y|x) \, dy = 1 \) for any positive \( \xi \) and \( \lim_{\xi \rightarrow +\infty} p^{\xi}_g(y|x) = \delta(y - g_\theta(x)) \). This approximation of the delta function by such a limit is practical due to the finite precision arithmetic used within computers; in other words, the delta function within computer arithmetic is actually an indicator function.

\newtheorem{prop}{Proposition}

\begin{restatable}{thm}{pr}
Let \( p_{\text{clean}}(y|x) > 0 \) be a finite and Lipschitz continuous density function with a unique global maximum and \( p^{\xi}_g(y|x) = \xi^n / 2^n \cdot \mathbf{1}_{y - g_\theta(x) \in [ - 1/\xi,  1/\xi]^n} \), then
\begin{equation}
   \lim_{\xi \rightarrow +\infty} \underset{g_\theta(x)}{\mathrm{arg\,min}} \ \chi^2_{\text{Pearson}} ( p_g^\xi || (p_{\text{clean}} + p_g^\xi) / 2 ) = \underset{y}{\mathrm{arg\,max}}\ p_{\text{clean}}(y|x) 
\end{equation}

\end{restatable}

Thus, LS-GAN training under ideal conditions should lead to the solution \( g_\theta(x) = \mathrm{arg\,max}_y\ p_{\text{clean}}(y|x) \) for the generator. In practice, however, success is highly dependent on technicalities, such as additional losses to stabilize training and architectures of neural networks. Below, we address these questions by revisiting the notion of perceptual loss for audio generation and assessing the effectiveness of neural architectures.

\section{Perceptual Loss for Speech Generation}
\label{section:loss_criteria}

Adversarial training is known for its instability issues~\citep{brock2018large}. It often leads to suboptimal solutions, mode collapse, and gradient explosions. For paired tasks, including speech enhancement, adversarial losses are often accompanied by additional regressive losses to stabilize training and guide the generator towards useful solutions \citep{kong2020hifi, su2020hifi}. In the context of GAN mode-seeking behaviour discussed above, regressive losses could be seen as a merit to push the generator towards the “right” (most-probable) mode. Therefore, finding an appropriate regression loss to guide adversarial training is of significant importance.

Historically, initial attempts to apply deep learning methods to speech enhancement were based on treating this problem as a predictive task \citep{defossez2020real, hao2021fullsubnet, chen2022fullsubnet+, isik2020poconet}. 
Following the principle of empirical risk minimization, the goal of predictive modelling is to find a model with minimal average error over the training data.
Given a noisy waveform or spectrogram, these approaches attempt to predict the clean signal by minimizing point-wise distance in waveform and spectrum domains or jointly in both domains, thus treating this problem as a predictive task.
However, given the severe degradations applied to the signal, there is an inherent uncertainty in the restoration of the speech signal (i.e., given the degraded signal, the clean signal is not restored unambiguously), which often leads to oversmoothing (averaging) of the predicted speech. A similar phenomenon is widely known in computer vision~\citep{ledig2017photo}.

One promising idea to reduce the averaging effect is to choose an appropriate representation space for regression, which is less “entangled” than waveform or spectrogram space. In simpler terms, the regression space should be designed so that averaged representation of sounds that are indistinguishable to humans (such as the same phoneme spoken by the same speaker with the same prosody) is still representation of this sound  (see \cref{app:motiv}).

We formulate two heuristic rules to compare different regression spaces based on their structure:

\begin{itemize}
    \item \textbf{Clustering rule:} Representations of identical speech sounds should form one cluster that is separable from clusters formed by different sounds.
    \item \textbf{SNR rule:} Representations of speech sounds contaminated by different levels of additive noise should move away from the cluster of clean sounds monotonically with the increase in the noise level.
\end{itemize}

The clustering rule ensures that minimizing the distance between samples in the feature space causes the samples to correspond to the same sound. The SNR rule ensures that minimizing the distance between features does not contaminate the signal with noise, meaning noisy samples are placed distantly from clean samples.

\begin{figure}[h!]
    \centering
    \includegraphics[height=5.5cm]{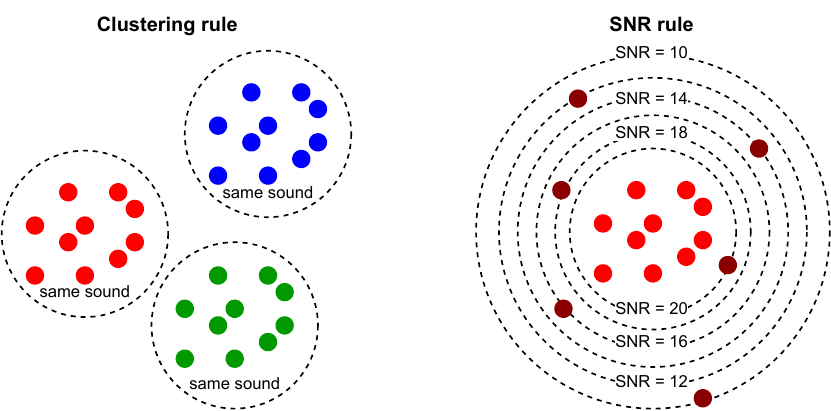}
    \caption{Illustration of heuristic rules for feature space structure. The Clustering rule (left) states that representations of the same speech sound should form clusters. The SNR rule (right) states that noise samples should deviate from the centre of the cluster as the amount of noise increases.
    Illustrations created using real samples are presented in ~\cref{fig:snr_real}, ~\cref{fig:clust_real}}
    \label{fig:rules}
\end{figure}

In practice, we check these conditions by the following procedure:
\begin{enumerate}
    \item We sample identical speech sounds with a multi-speaker VITS text-to-speech model~\citep{kim2021conditional}. We define the group of waveforms corresponding to the same sound as 0.5-second waveform segments generated by the same speaker with the same text and the same phoneme duration (i.e., the stochastic duration predictor is used only once for each group). Thus, the waveform variation is produced due to sampling of latents from the prior distribution. We note that we do not explicitly fix the prosody; however, we observed that prosody variation was low by default and the sounds generated by this procedure were mostly perceptually indistinguishable from each other while corresponding to different waveform realizations. Overall, we define 354 groups of sounds with 80 speakers saying 177 different phrases, 2 different speakers for each phrase; we sample 20 samples for each group, totalling 7080 waveform segments corresponding to 354 groups.
    \item Each waveform segment is then mapped to the feature space, and the resulting tensors are flattened to obtain a vector for each sample. The vectors are clustered using K-means clustering~\citep{macqueen1967some}, the number of clusters is set to the number of groups (354). After clustering, we compute the Rand index~\citep{rand1971objective} of the clustering produced by K-means with the clustering induced by initial groups of sounds splits. We treat the resulting number as a way to measure the adherence to the \textbf{clustering rule}.
    \item Each waveform segment within a group is randomly mixed with noise at SNR levels ranging from 10 to 20 dB. The segments are then mapped to the feature space to obtain a vector for each sample. For each noisy sample, we compute the Euclidean distance between its features and the centre of the clean feature cluster. After that, we compute the negative Spearman's rank correlation between the SNR level and the distance from the sample to the centre of the cluster. The negative correlations are averaged over all groups of sounds, and the resulting number is treated as a quantitative measure of adherence to the \textbf{SNR rule}.
    
\end{enumerate}

Using these metrics, we assess the effectiveness of different feature spaces formed by several speech feature extractors, as well as conventional representations of speech. Namely, we produce features by Wav2Vec 2.0~\citep{baevski2020wav2vec}, WavLM~\citep{chen2022wavlm}, the encoder of EnCodec~\citep{defossez2023high}, and CDPAM~\citep{manocha2021cdpam}. As a conventional representation, we use waveform and spectrogram features. For Wav2Vec 2.0 and WavLM, we consider the output of the last transformer layer and the output of the convolutional encoder as separate cases. We also train a HiFi-GAN generator \citep{kong2020hifi} with each feature type used as a representation for mean squared error loss computation on a neural vocoding task. To assess experimentally the suitability of the feature space to be used as a loss function for waveform generation, we report MOS scores for samples generated by vocoders trained with each feature map. The results are presented in \cref{table:feature}.

\begin{table}[!h]
    \centering
    \caption{Comparison of different features using Clustering rule, SNR rule, and MOS on neural vocoding.}
    \label{table:feature}
    \begin{tabular}{l c c c H}
    \toprule
    \mlcell{Feature \\ space} & \mlcell{Rand score ($\uparrow$) \\(Clustering rule)}  &  \mlcell{Negative correlation ($\uparrow$) \\ (SNR rule)}  & \mlcell{MOS ($\uparrow$) \\ (Vocoding)} & GMAC \\
    \midrule
    Waveform & \(0.00 \pm 0.00\) & \(0.31 \pm 0.02\) & Failed & 0.29 \\
    Spectrogram & \(0.00 \pm 0.00\) & \(0.08 \pm 0.03\) & \(1.78 \pm 0.08\) & 0.21 \\
    Wav2Vec 2.0 & \(0.25 \pm 0.03\) & \(0.19 \pm 0.03\) & \(1.65 \pm 0.08\) & 0.21 \\
    Wav2Vec 2.0-conv & \(0.94 \pm 0.01\) & \(0.78 \pm 0.02\) & \(2.23 \pm 0.09\) & 0.21 \\
    WavLM & \(0.46 \pm 0.05\) & \(0.46 \pm 0.03\) & \(1.71 \pm 0.07\) & 0.21 \\
    WavLM-conv & \(\mathbf{0.96 \pm 0.01}\) & \(\mathbf{0.89 \pm 0.02}\) & \(\mathbf{3.27 \pm 0.10}\) & 0.21 \\
    EnCodec & \(0.55 \pm 0.03\) & \(0.67 \pm 0.03\) & \(1.80 \pm 0.08\) & 0.21 \\
    CDPAM & \(0.00 \pm 0.00\) & \(0.17 \pm 0.03\) & Failed & 0.21 \\
    \bottomrule
    \end{tabular}
\end{table}

As can be seen, the features extracted by the convolutional encoder of the WavLM model present the best performance according to both our internal metrics and MOS score on the neural vocoding task. We have also tested the other layers of WavLM but did not observe significant improvements when using other layers (see \cref{app:wavlm_l}). Empirically, we have found that if WavLM-conv feature loss is used together with spectrogram loss (L1-distance magnitudes of STFT), the performance on the vocoding task increases significantly, likely due to the looseness of the WavLM representation (\cref{app:spec_loss}). Without any adversarial training, the WavLM-conv+L1-STFT loss (LMOS-loss, \cref{lmos}) achieves an MOS score of \(4.31 \pm 0.08\), approaching the original adversarially trained HiFi GAN generator which achieves \(4.69 \pm 0.05\) (\cref{app:vocod}).

\section{FINALLY}


\subsection{Architecture}

Our method is based on HiFi++ \citep{andreev2022hifi++}. The HiFi++ generator is a four-component neural network consisting of SpectralUNet, Upsampler, WaveUNet, and SpectralMaskNet modules. 
SpectralUNet is responsible for initial preprocessing of audio in the spectral domain using two-dimensional convolutions.
The Upsampler is a HiFi-GAN generator-based module that increases the temporal resolution of the input tensor, mapping it to the waveform domain.
WaveUNet performs post-processing in the waveform domain and improves the output of the Upsampler by incorporating phase information gleaned directly from the raw input waveform.
Finally, SpectralMaskNet is applied to perform spectrum-based post-processing and, thus, remove any possible artefacts that remained after WaveUNet.
Thus, the model alternates between time and frequency domains, allowing for effective audio restoration.

We introduce two modifications to the HiFi++ generator's architecture. 
First, we modify the generator by incorporating WavLM-large model output (last hidden state of the transformer) as an additional input to the Upsampler. Prior works~\citep{hung2022boosting, byun2023empirical} have demonstrated the usefulness of Self-Supervised Learning (SSL) features for speech enhancement tasks, and we validate this by observing significant performance gains from using SSL features. Second, we introduce the Upsample WaveUNet at the end of the generator. 
This module acts as a learnable upsampler of the signal sampling rate. For its architecture, we use the WaveUNet with an additional convolutional upsampling block in the decoder that upsamples the temporal resolution by 3 times. This allows the model to output a 48 kHz signal while taking a 16 kHz signal as input.

\begin{figure}[ht!]
        \centering
        \includegraphics[height=4.0cm]{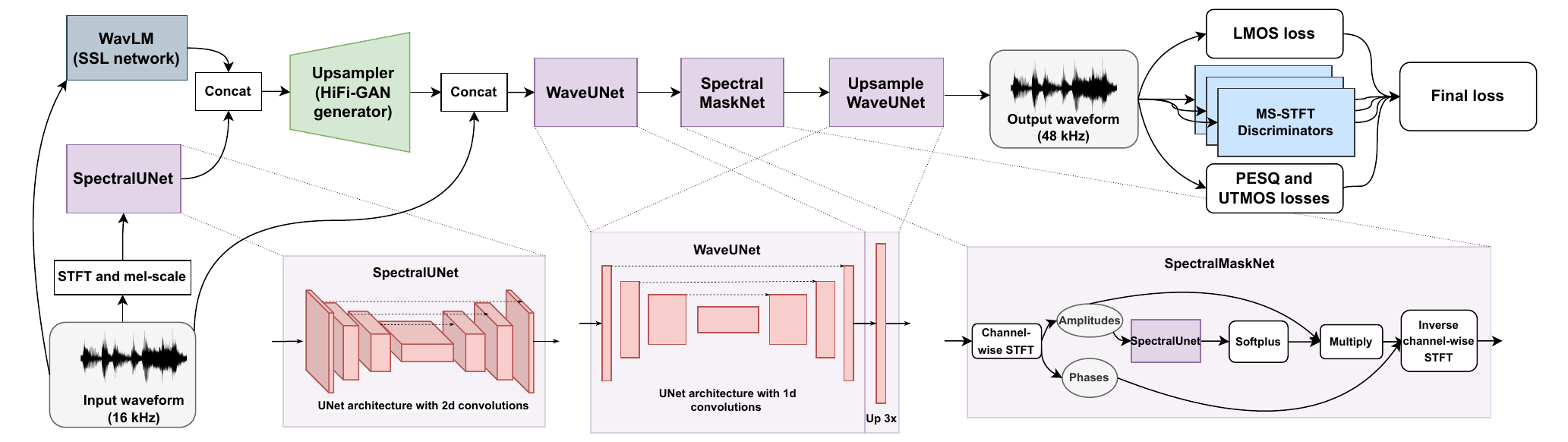}
        \caption{FINALLY model architecture.}
        \label{fig:arch}
              \vspace{-0.4cm}

\end{figure}
\subsection{Data and training}

We use LibriTTS-R~\citep{koizumi2023libritts}, DAPS-clean~\citep{mysore2014can} as the sources of clean speech data. LibriTTS-R is used at 16 kHz, while DAPS at 48 kHz.
Noise samples were taken from the DNS dataset~\citep{dubey2022icassp}. After mixing with noise, we apply several digital distortion effects (see \cref{app:augs} for details). 

We train the model in three stages. The first two stages concentrate on restoring the original speech content, and the final stage aims to enhance the aesthetic perception of the speech. The multi-stage approach is necessary due to the characteristics of the employed datasets: the LibriTTS-R dataset has a lot of samples but limited perceptual quality, whereas the DAPS dataset is high-quality but contains a smaller number of samples. Consequently, we utilize the LibriTTS-R dataset for learning speech content restoration and the DAPS dataset for aesthetic stylization.

The loss functions that we use can be written as follows:

\begin{equation}
        \mathcal{L}_{\text{LMOS}}(\theta) =  \underset{x,y \sim p(x, y)}{\mathbb{E}} \left [ 100 \cdot\|\phi(y) - \phi(g_\theta(x))\|_2^2 + \||\mathrm{STFT}(y)| - |\mathrm{STFT}(g_\theta(x))|\|_1 \right ], \label{lmos}
\end{equation}

 \vspace{-0.4cm}

\begin{align}
    &\mathcal{L}_{\text{gen}}(\theta) = \underbrace{\overbrace{\underbrace{\lambda_{\text{LMOS}}  \cdot \mathcal{L}_{\text{LMOS}}(\theta)}_{\text{1st stage (16 kHz)}} + \lambda_{\text{GAN}} \cdot \mathcal{L}_{\text{GAN-gen}}(\theta) + \lambda_{\text{FM}} \cdot \mathcal{L}_{\text{FM}}(\theta)}^{\text{2nd stage (16 kHz)}}  + \lambda_{\text{HF}} \cdot \mathcal{L}_{\text{HF}}(\theta)}_{\text{3rd stage (48 kHz)}}, \\
    & \mathcal{L}_{\text{disc}}(\varphi_i) = \mathcal{L}_{\text{GAN-disc}}(\varphi_i), \quad i = 1, \ldots, k.
\end{align}

Here, \(\phi\) denotes the WavLM-Conv feature mapping, \(g_\theta(x)\) denotes the generator neural network with parameters \(\theta\), \(\mathcal{L}_{\text{GAN-gen}}(\theta)\) denotes the LS-GAN generator loss~\citep{mao2017least}, \(\mathcal{L}(\theta)\) denotes the combined generator loss, \(\mathcal{L}_{\text{GAN-disc}}(\varphi_i)\) denotes the LS-GAN discriminator~\citep{mao2017least} loss for the \(i\)-th discriminator with parameters \(\varphi_i\), \(\mathcal{L}_{\text{FM}}\) denotes the feature matching loss~\citep{kumar2019melgan, defossez2023high}, \(\mathcal{L}_{\text{HF}}\) denotes the human feedback loss, and \(\lambda_{\textbf{*}}\) denotes the corresponding loss weights. Following \cite{defossez2023high}, we employ 5 discriminators with STFT window lengths of \([ 2048, 1024, 512, 256, 128]\) for the 2nd stage and 5 discriminators with STFT window lengths of \([4096, 2048, 1024, 512, 256]\) for the 3rd stage, which allows for the capture of spectral information at different resolutions.

\textbf{Stage 1}: Firstly, we train the FINALLY architecture without Upsample WaveUNet (we refer to this truncated architecture as FINALLY-16) at a 16 kHz sampling rate on the LibriTTS-R dataset. We train the model with the proposed LMOS regression loss to provide the generator with a better initialization before adversarial training.

\textbf{Stage 2}: Second, we start adversarial training of FINALLY-16 with MS-STFT discriminators~\citep{defossez2023high} and the LMOS loss. The learning of the generator undergoes a linear warm-up to give the discriminators time to learn meaningful representations. At this stage, we focus the generator on producing a reliable reconstruction of linguistic content by assigning larger values for LMOS and feature matching losses (\(\lambda_{\text{GAN-gen}} = 0.4\), \(\lambda_{\text{FM}} = 20\), \(\lambda_{\text{LMOS}} = 20\)).

\textbf{Stage 3}: Lastly, we attach Upsample WaveUNet to FINALLY-16 and start adversarial training of the FINALLY model to produce 48 kHz output. We subsample 48 kHz waveforms of the DAPS dataset and apply distortions to form the 16 kHz input. The discriminators are initialized randomly, and the learning rate for the generator undergoes a warm-up similar to the second stage. This stage is focused on producing the final output; therefore, we focus the generator on perceptual quality by increasing the relative weight of GAN loss (\(\lambda_{\text{GAN-gen}} = 5\), \(\lambda_{\text{FM}} = 15\), \(\lambda_{\text{LMOS}} = 0.5\)) and introducing additional human feedback loss \(\mathcal{L}_{\text{HF}}\), which is based on UTMOS and PESQ metrics. The UTMOS loss \citep{saeki2022utmos} is based on a neural network model that simulates subjective MOS metric results. On the other hand, the PESQ loss\footnote{\href{https://github.com/audiolabs/torch-pesq}{https://github.com/audiolabs/torch-pesq}} delivers a differentiable version of the PESQ metric \citep{rix2001perceptual}, as presented by \citet{kim2019end, martin2018deep}. Both UTMOS and PESQ help enhance speech aesthetic quality, as they incorporate insights from human preference studies. These metrics are differentiable with respect to their inputs, making them suitable for use as loss functions (multiplied by negative constants) \(\mathcal{L}_{\text{HF}} = -20 \cdot \mathcal{L}_{\text{UTMOS}} - 2 \cdot \mathcal{L}_{\text{PESQ}}\), \(\lambda_{\text{HF}} = 1\).

\section{Related Work}

\paragraph{Self-supervised features for speech enhancement}

Several works have employed self-supervised (SSL) features ~\citep{baevski2020wav2vec, chen2022wavlm, hsu2021hubert} for training of speech enhancement models as intermediate representation~\citep{wang2024selm, koizumi2023miipher}, auxiliary space for loss computation~\citep{sato2023downstream, hsieh2020improving, close2023effect, close2023perceive} or input features~\citep{hung2022boosting, byun2023empirical}. 
\cite{sato2023downstream, close2023effect, close2023perceive, hsieh2020improving} proposed to use features of self-supervised models as an auxiliary space for regression loss computation. 
In our work, we similarly study the effect of SSL features on speech enhancement models; however, our study systematically compares different feature backbones and develops criteria for probing the structure of different feature spaces. 
In particular, we show that features of the WavLM's convolutional encoder are the most effective for the loss function, while recent work~\citep{sato2023downstream} has used the outputs of transformer layers. 
\citet{close2023effect, close2023perceive} proposed a similar loss function for speech enhancement based on convolutional features of HuBERT~\citep{hsu2021hubert}, our work can be considered as extension of this work as we show that additional spectrogram loss greatly benefits the quality and provide experimental evidence for the advantages of LMOS-loss usage for adversarial training.  

\paragraph{GAN-based approaches}

The HiFi GAN works~\citep{su2020hifi, su2021hifi} consider a GAN-based approach to speech enhancement, which is similar to ours. 
Importantly, these works base their generator architectures on feed-forward WaveNet~\citep{rethage2018wavenet}, a fully-convolutional neural network that operates at full input resolution, leading to slow training and inference times. We show that our model is able to achieve superior quality compared to this model while being much more efficient.

\cite{koizumi2023miipher} proposes to use w2v-BERT features~\citep{chung2021w2v} as intermediate representations for speech restoration. The features extracted from the noisy waveform are processed by a feature cleanser which is conditioned on the text representation extracted from transcripts via PnG-BERT, and speaker embedding extracted from the waveform. 
The feature cleanser is trained to minimize a regression loss between w2v-BERT features extracted from the clean signal and the predicted ones. At the second stage, the WaveFit vocoder is trained to synthesize a waveform based on the predicted features~\citep{koizumi2023wavefit}. 
An important difference with our work is that the method uses a text transcript, while our model does not. Despite this difference, we show that our model delivers better perceptual quality as reported by human listeners (\cref{app:miip}).

\paragraph{Diffusion-based approaches}

The recent success of diffusion-based generative models has led to their use in a wide array of applications, including speech enhancement. Numerous studies~\citep{lemercier2023storm, welker2022speech, richter2023speech, lay2023reducing} have applied diffusion models in various configurations to generate high-fidelity, clear speech from noisy input. 
For instance, \cite{welker2022speech, richter2023speech} introduced a novel stochastic diffusion process to design a generative
model for speech enhancement in the complex STFT domain for speech denoising and dereverberation. In the UNIVERSE study~\citep{serra2022universal}, the authors propose a diffusion model for universal speech enhancement. They create a paired dataset using 55 different distortions and train a conditional diffusion model on it. Although their model performs well in terms of quality, it requires up to 50 diffusion steps to produce the final audio. The authors demonstrate that the number of steps can be reduced; however, the effect of this reduction on perceptual quality remains somewhat unclear. In our experiments, we show that our model can achieve similar results with just a single forward pass.

\section{Results}

\subsection{Evaluation} 

We evaluate our models using the following sources of data:

\textbf{VoxCeleb Data}: 50 audio clips selected from VoxCeleb1~\citep{nagrani2017voxceleb} to cover the Speech Transmission Index (STI) range of 0.75-0.99 uniformly and balanced across male and female speakers.

\textbf{UNIVERSE Data}: 100 audio clips randomly generated by the UNIVERSE~\citep{serra2022universal} authors from clean utterances sampled from VCTK and Harvard sentences, together with noises/backgrounds from DEMAND and FSDnoisy18k. The data contains various artificially simulated distortions including band limiting, reverberation, codec, and transmission artefacts. Please refer to~\citep{serra2022universal} for further details. The validation data of UNIVERSE is artificially simulated from clean speech recordings using the same pipeline that the authors utilized for training. Therefore, we must note that the comparison is conducted in a manner advantageous to UNIVERSE, since our data simulation pipeline is different.

\textbf{VCTK-DEMAND}: we use validation samples from popular Valentini denoising benchmark~\citep{valentini2017noisy}. This dataset is used for a broad comparison with a wide range of speech enhancement models. The test set (824 utterances) includes artificially simulated noisy samples from 2 speakers with 4 SNR (17.5, 12.5, 7.5, and 2.5 dB). 

We utilize DNSMOS~\citep{reddy2022dnsmos}, UTMOS~\citep{saeki2022utmos}, and WV-MOS~\citep{andreev2022hifi++} as non-intrusive metrics to objectively assess the samples generated by our speech enhancement model on the VoxCeleb dataset. The non-intrusive nature of these metrics is essential since the dataset comprises recordings from real-life scenarios, lacking ground-truth samples. 

In addition, we compute the Phoneme Error Rate (PhER) and Word Error Rate (WER) by comparing ground truths with the generated samples (see \cref{app:eval} for details) for both the UNIVERSE and VCTK-DEMAND datasets. For subjective quality assessment, we conduct 5-point Mean Opinion Score (MOS) tests. All audio clips are normalized to ensure volume differences do not influence the raters' evaluations. The raters are required to be English speakers using appropriate listening equipment (more details in \cref{app:subj}). 

The Real-Time Factor (RTF) is determined by measuring the processing time (in seconds) for a 10-second audio segment on a V100 GPU and then dividing this time by 10. All confidence intervals are calculated using bootstrapping method.

We also evaluate our model on the VCTK-DEMAND dataset~\citep{valentini2017noisy} with additional metrics such as PESQ~\citep{pesq-metric}, STOI~\citep{stoi-metric}, and SI-SDR~\citep{si-sdr-metric}. These metrics are included to ensure consistent comparison with previous works.

\subsection{Comparison with existing approaches}

We consider BBED~\citep{lay2023reducing}, STORM~\citep{lemercier2023storm}, and UNIVERSE~\citep{serra2022universal} diffusion models, along with Voicefixer and DEMUCS regression models, as our baselines. In addition, we consider our closest competitor, HiFi-GAN-2, as a GAN-based baseline. The data for comparison with HiFi-GAN-2 and UNIVERSE were taken from their demo pages, since the authors did not release any code. We conduct comparisons with BBED, STORM, Voicefixer, DEMUCS, and HiFi-GAN-2 on real-world VoxCeleb1 samples and the comparison with UNIVERSE on the simulated data, provided by the authors of this work. The results are presented in Table~\ref{table:combined}. We also compare these models on VCTK-DEMAND dataset, results can be found  in Table~\ref{tab:model_comparison}. We complement this table by two additional models: MetricGAN+~\citep{MetricGAN+} and DB-AIAT~\citep{DB-AIAT}. 

\begin{table*}[!h]
\centering
\caption{Comparison with prior work on Voxceleb and UNIVERSE validation data.}
\label{table:combined}
\scalebox{0.9}{
\begin{tabular}{ llllll|ll}
    \toprule
    \multicolumn{7}{c}{\textbf{VoxCeleb (HiFi-GAN-2 validation set, real data)}} \\
    \midrule
    Model & MOS ($\uparrow$)  & UTMOS ($\uparrow$)  & WV-MOS ($\uparrow$)  & DNSMOS ($\uparrow$)  & - & RTF ($\downarrow$)  \\
    \midrule
    Input & 3.46 $\pm$ 0.07 & 2.76 $\pm$ 0.13 & 2.90 $\pm$ 0.16 & 2.72 $\pm$ 0.11 & - & - \\
    VoiceFixer & 3.41 $\pm$ 0.07 & 2.60 $\pm$ 0.09 & 2.79 $\pm$ 0.09 & 3.08 $\pm$ 0.06 & - & 0.02 \\
    DEMUCS & 3.79 $\pm$ 0.07 & 3.51 $\pm$ 0.08 & 3.72 $\pm$ 0.08 & 3.27 $\pm$ 0.04 & - & 0.08 \\
    STORM & 3.75 $\pm$ 0.06 & 3.29 $\pm$ 0.08 & 3.54 $\pm$ 0.09 & 3.17 $\pm$ 0.04 & - & 1.05 \\
    BBED & 3.97 $\pm$ 0.06 & 3.30 $\pm$ 0.10 & 3.47 $\pm$ 0.08 & 3.23 $\pm$ 0.04 & - & 0.43 \\
    HiFi-GAN-2 & 4.47 $\pm$ 0.05 & 3.67 $\pm$ 0.09 & \textbf{3.96 $\pmb{\pm}$ 0.06} & \textbf{3.32 $\pmb{\pm}$ 0.03} & - & 0.50 \\
    Ours & \textbf{4.63 $\pmb{\pm}$ 0.04} & \textbf{4.05 $\pmb{\pm}$ 0.07} & \textbf{3.98 $\pmb{\pm}$ 0.06} & \textbf{3.31 $\pmb{\pm}$ 0.04} & - & \textbf{0.03} \\
    \midrule
    \multicolumn{7}{c}{\textbf{UNIVERSE validation set (simulated data)}} \\
    \midrule
    Model & MOS ($\uparrow$)  & UTMOS ($\uparrow$)  & WV-MOS ($\uparrow$)  & DNSMOS ($\uparrow$)  & PhER ($\downarrow$)  & RTF ($\downarrow$)  \\
    \midrule
    Input & 2.87 $\pm$ 0.05 & 2.27 $\pm$ 0.28 & 1.72 $\pm$ 0.61 & 2.25 $\pm$ 0.19 & 0.31 $\pm$ 0.05 & - \\
    Ground Truth & 4.39 $\pm$ 0.05 & 4.26 $\pm$ 0.06 & 4.28 $\pm$ 0.06 & 3.33 $\pm$ 0.04 & 0 & - \\
    UNIVERSE & 4.10 $\pm$ 0.07 & 3.89 $\pm$ 0.15 & 3.85 $\pm$ 0.12 & \textbf{3.23 $\pmb{\pm}$ 0.07} & 0.20 $\pm$ 0.04 & 0.5 \\
    Ours (16 kHz) & 3.99 $\pm$ 0.07 & \textbf{4.21 $\pmb{\pm}$ 0.10} & \textbf{4.43 $\pmb{\pm}$ 0.07} & \textbf{3.25 $\pmb{\pm}$ 0.05} & \textbf{0.14 $\pmb{\pm}$ 0.03} & \textbf{0.03} \\
    Ours & \textbf{4.23 $\pmb{\pm}$ 0.07} & \textbf{4.21 $\pmb{\pm}$ 0.10} & \textbf{4.43 $\pmb{\pm}$ 0.08} & \textbf{3.25 $\pmb{\pm}$ 0.05} & \textbf{0.14 $\pmb{\pm}$ 0.03} & \textbf{0.03} \\
    \bottomrule
\end{tabular}}
\end{table*}

\begin{table*}[h!]
\centering
\caption{Comparison with the baselines on VCTK-DEMAND.}
\label{tab:model_comparison}
\scalebox{0.69}{
\begin{tabular}{lcccccccc}
\toprule
Model                        & MOS ($\uparrow$)        & UTMOS ($\uparrow$)      & WV-MOS ($\uparrow$)      & DNSMOS ($\uparrow$)     & PESQ  ($\uparrow$)       & STOI ($\uparrow$)        & SI-SDR ($\uparrow$)      & WER ($\downarrow$)      \\
\toprule
Input                        & $3.18 \pm 0.07$ & $3.06 \pm 0.14$ & $2.99 \pm 0.24$ & $2.53 \pm 0.10$ & $1.98 \pm 0.17$  & $0.92 \pm 0.01$ & $8.4 \pm 1.2$   & $0.09 \pm 0.03$ \\
\midrule
MetricGAN+                   & $3.75 \pm 0.06$ & $3.62 \pm 0.09$ & $3.89 \pm 0.10$ & $2.95 \pm 0.05$ & $3.14 \pm 0.10$  & $0.93 \pm 0.01$ & $8.6 \pm 0.7$   & $0.10 \pm 0.04$ \\
DEMUCS                       & $3.95 \pm 0.06$ & $3.95 \pm 0.05$ & $4.37 \pm 0.06$ & $3.14 \pm 0.04$ & $3.04 \pm 0.12$  & \textbf{0.95 $\pmb{\pm}$ 0.01} & $18.5 \pm 0.6$  & \textbf{0.07 $\pmb{\pm}$ 0.03} \\
HiFi++                       & $4.08 \pm 0.05$ & $3.89 \pm 0.06$ & $4.36 \pm 0.06$ & $3.10 \pm 0.04$ & $2.90 \pm 0.12$  & \textbf{0.95 $\pmb{\pm}$ 0.01} & $17.9 \pm 0.6$  & $0.08 \pm 0.03$ \\
HiFi-GAN-2                   & $4.13 \pm 0.05$ & $3.99 \pm 0.05$ & $4.26 \pm 0.05$ & $3.12 \pm 0.05$ & $3.14 \pm 0.12$  & \textbf{0.95 $\pmb{\pm}$ 0.01} & $18.6 \pm 0.6$  & \textbf{0.07 $\pmb{\pm}$ 0.03} \\
DB-AIAT                      & $4.22 \pm 0.05$ & $4.02 \pm 0.05$ & $4.38 \pm 0.06$ & $3.18 \pm 0.04$ & \textbf{3.26 $\pmb{\pm}$ 0.12}  & \textbf{0.96 $\pmb{\pm}$ 0.01} & \textbf{19.3 $\pmb{\pm}$ 0.8}  & \textbf{0.07 $\pmb{\pm}$ 0.03} \\
Ours (16 kHz)                & \textbf{4.41 $\pmb{\pm}$ 0.04} & \textbf{4.32 $\pmb{\pm}$ 0.02} & \textbf{4.87 $\pmb{\pm}$ 0.05} & \textbf{3.22 $\pmb{\pm}$ 0.04} & $2.94 \pm 0.10$  & $0.92 \pm 0.01$ & $4.6 \pm 0.3$   & \textbf{0.07 $\pmb{\pm}$ 0.03} \\
Ours (48 kHz)                & \textbf{4.66 $\pmb{\pm}$ 0.04} & \textbf{4.32 $\pmb{\pm}$ 0.02} & \textbf{4.87 $\pmb{\pm}$ 0.05} & \textbf{3.22 $\pmb{\pm}$ 0.04} & $2.94 \pm 0.10$  & $0.92 \pm 0.01$ & $4.6 \pm 0.3$   & \textbf{0.07 $\pmb{\pm}$ 0.03} \\
\midrule
GT (16 kHz)        & $4.26 \pm 0.05$ & $4.07 \pm 0.04$ & $4.52 \pm 0.04$ & $3.16 \pm 0.04$ & --               & --              & --              & --               \\
GT (48 kHz)        & $4.56 \pm 0.03$ & $4.07 \pm 0.04$ & $4.52 \pm 0.04$ & $3.16 \pm 0.04$ & --               & --              & --              & --               \\
\bottomrule
\end{tabular}}
\end{table*}

Importantly, our study confirms the observation from \cite{serra2022universal} that speech enhancers based on generative models significantly outperform regression-based approaches. Our model performs comparably in terms of perceptual MOS quality to all the considered baselines, while being more than five times as efficient as the closest competitors, HiFi-GAN-2 and UNIVERSE. We have also found that our model is less prone to hallucinating linguistic content than UNIVERSE, delivering a lower Phoneme Error Rate (PhER) value.

Lastly, we would like to comment on the results in \cref{tab:model_comparison}. Our model outperforms baselines in subjective evaluation and no-reference metrics, e.g. UTMOS, but underperforms in terms of PESQ~\citep{pesq-metric}, STOI~\citep{stoi-metric} and SI-SDR~\citep{si-sdr-metric}. Notably, there are numerous works consistently reporting low correlation of reference-based metrics with human perceptual judgment~\citep{manocha2022sim-is-bad, Manjunath-limits, andreev2022hifi++}. In particular, the study~\citep{manocha2022sim-is-bad} reports that no-reference metrics (including DNSMOS, reported in our work) correlate significantly better with human perception and therefore have higher relevance for objective comparison between methods. Furthermore, in our study, we report the MOS score, which directly reflects human judgments of restoration quality.

\subsection{Ablation study}
To validate the improvements proposed in this work, we conduct an ablation study assessing the effectiveness of the design choices made.

Firstly, we compare the LMOS loss against two other regression losses in the context of training a small speech enhancement model (10 times smaller than the final model; please see the \cref{app:ablation_details} for details). The first regression loss is the Mel-Spectrogram loss, which was proposed by \citet{kong2020hifi}. As the second alternative, we employ the Reconstruction loss (RecLoss) proposed by \citet{defossez2023high}. It consists of a combination of L1 and L2 distances between mel-spectrograms computed with different resolutions. For these experiments, we conducted a grid search for the weights of each reconstruction loss (details are in  \cref{app:ablation_details}) and report results for the best option in \cref{table:loss_ablations}.

\begin{table}[!h]
    \caption{Ablation study (VoxCeleb real data).}
    \label{table:loss_ablations}
    \centering
    \begin{tabular}{c l c c c c }
        \toprule
        & \mlcell{Loss} & \mlcell{MOS ($\uparrow$)}   &  \mlcell{UTMOS ($\uparrow$)}   & \mlcell{WV-MOS ($\uparrow$)} & \mlcell{DNSMOS ($\uparrow$)} \\
       
        \midrule
        & input                          & \(3.46 \pm 0.07\) & \(2.77 \pm 0.14\) & \(2.88 \pm 0.16\) &  \(2.72 \pm 0.11\)  \\
        
        \midrule
        \multirow{7}{*}[0em]{\rotatebox[origin=c]{90}{2nd stage}} 
        & w/o reg. loss         & \(3.71 \pm 0.08\) & \(3.05 \pm 0.09\) & \(2.70 \pm 0.11\) &  \(3.18 \pm 0.06\)  \\
        & w/ L1Spec                          & \(4.15 \pm 0.06\) & \(3.45 \pm 0.08\) & \(3.41 \pm 0.07\) &  \(\textbf{3.28} \pmb{\pm} \textbf{0.04}\)   \\
        & w/ RecLoss                         & \(4.07 \pm 0.06\) & \(3.46 \pm 0.07\) & \(3.43 \pm 0.06\) &  \(\textbf{3.28} \pmb{\pm} \textbf{0.04}\)   \\
        & w/ LMOS        & \(4.20 \pm 0.05\) & \(3.48 \pm 0.08\) & \(3.58 \pm 0.06\) & \( 3.26 \pm 0.04 \)    \\
        \cmidrule(lr){2-6}
        & + WavLM enc.   & \(4.21 \pm 0.06\) & \(3.65 \pm 0.08\) & \(3.75 \pm 0.05\) &  \(3.26 \pm 0.04\)  \\
        \cmidrule(lr){2-6}
        & + Scaling   & \(4.30 \pm 0.05\) & \(3.83 \pm 0.07\) & \(\textbf{4.00} \pmb{\pm} \textbf{0.06}\) &  \(3.21 \pm 0.05\)  \\
        
        \midrule
        \multirow{2}{*}[0em]{\rotatebox[origin=c]{90}{3rd st.}}  & + 3rd stage   & \(4.59 \pm 0.05\) & \(3.78 \pm 0.07\) & \(\textbf{3.99} \pmb{\pm} \textbf{0.06}\) &  \(\textbf{3.29} \pmb{\pm} \textbf{0.04}\)  \\
        \cmidrule(lr){2-6}
        & + HF Loss & \(\textbf{4.63} \pmb{\pm} \textbf{0.04}\) & \(\textbf{4.05} \pm \textbf{0.07}\) & \(\textbf{3.98} \pmb{\pm} \textbf{0.06}\) &  \(\textbf{3.31} \pmb{\pm} \textbf{0.04}\)  \\
        \bottomrule
    \end{tabular}
\end{table}

The other proposed improvements bring incremental gains in perceptual MOS quality. Firstly, we add the features of the WavLM encoder as an additional input to HiFi++. Next, we scale the architecture by increasing the number of channels within the model. After that, we concatenate the Upsample WaveUNet with the FINALLY-16 architecture and implement the 3rd stage of training on the DAPS-clean dataset (48 kHz). Finally, we use additional human feedback losses (HF Loss) during the 3rd stage to further improve perceptual quality.

Since \cref{table:loss_ablations} does not clearly demonstrate the advantages of using the WavLM \citep{chen2022wavlm} encoder in terms of MOS, we provide additional ablation study on more challenging UNIVERSE~\citep{serra2022universal} validation dataset. The results of these additional experiments, presented in \cref{tab:wavlm_ablation}, clearly demonstrate the importance of WavLM encoder.

\begin{table}[ht]
\centering
\caption{Ablation of WavLM encoder on UNIVERSE validation data.}
\begin{tabular}{lccccc}
\toprule
 & MOS ($\uparrow$) & UTMOS ($\uparrow$) & WV-MOS ($\uparrow$) & DNSMOS ($\uparrow$) & PhER ($\downarrow$) \\
\midrule
w/o WavLM & $3.49 \pm 0.08$ & $3.33 \pm 0.18$ & $3.80 \pm 0.15$ &  \(\textbf{3.15} \pmb{\pm} \textbf{0.09}\)& $0.27 \pm 0.04$ \\
w/ WavLM &  \(\textbf{3.75} \pmb{\pm} \textbf{0.07}\) & \(\textbf{3.56} \pmb{\pm} \textbf{0.20}\) & \(\textbf{3.99} \pmb{\pm} \textbf{0.08}\) & \(\textbf{3.07} \pmb{\pm} \textbf{0.16}\) & \(\textbf{0.21} \pmb{\pm} \textbf{0.04}\)\\
\bottomrule
\end{tabular}
\label{tab:wavlm_ablation}
\end{table}

\section{Conclusion}

In conclusion, we theoretically demonstrate that LS-GAN training encourages the selection of the point of maximum density within the conditional clean speech distribution, aligning naturally with the objectives of speech enhancement. Our empirical investigation identifies WavLM as an effective backbone for perceptual loss, supporting adversarial training. By integrating WavLM-based perceptual loss into the MS-STFT adversarial training pipeline and enhancing the HiFi++ architecture with a WavLM encoder, we develop a novel speech enhancement model, FINALLY, which achieves state-of-the-art performance, producing clear and high-quality speech at 48 kHz.


%

\bibliography{A}
\bibliographystyle{icml2022}

\newpage
\appendix

 \section{Proof}

\pr*

\begin{proof}
\begin{align*}
& \chi^2_{\text{Pearson}} ( p_g^\xi  (p_{\text{clean}} + p_g^\xi) / 2 ) =  \int \frac{(p_g^\xi(y|x) - (p_{\text{clean}}(y|x) + p_g^\xi(y|x)) / 2)^2}{(p_{\text{clean}}(y|x) + p_g^\xi(y|x)) / 2} dy \\
& = \int \limits_{y - g_\theta(x) \in [- 1/\xi, 1/\xi]^n} \frac{1}{2} p_g^\xi(y|x) \frac{(1 - p_{\text{clean}}(y|x)/p_g^\xi(y|x))^2}{1 + p_{\text{clean}}(y|x) / p_g^\xi(y|x)} dy \\
& +  \int \limits_{\mathbb{R}^n  \setminus \{y - g_\theta(x) \in [- 1/\xi, 1/\xi]^n\}} \frac{1}{2} \frac{(p_{\text{clean}}(y|x))^2}{p_{\text{clean}}(y|x)} dy  \\
& = \int \limits_{y - g_\theta(x) \in [- 1/\xi, 1/\xi]^n} \frac{\xi^n}{2^{n+1}} \frac{(1 - 2^n p_{\text{clean}}(y|x)/\xi^n)^2}{1 + 2^n p_{\text{clean}}(y|x) / \xi^n} dy \\
& + \int \limits_{\mathbb{R}^n  \setminus \{y - g_\theta(x) \in [- 1/\xi, 1/\xi]^n\} } \frac{1}{2} p_{\text{clean}}(y|x) dy .\\
\end{align*}

Since \( p_{\text{clean}}(y|x) \) is Lipschitz continuous, there exist \( L > 0 \) and \( C > 0 \) such that for any \( \xi > C \)  if \( y - g_\theta(x)\in [- 1/\xi, 1/\xi]^n \), then \( p_{\text{clean}}(y|x) \in (p_{\text{clean}}(g_\theta(x)|x) - L / \xi, p_{\text{clean}}(g_\theta(x)|x)+  L / \xi) \).

Therefore, the divergence could be lower-bounded by

\begin{align*}
\frac{1}{2} \cdot \frac{(1 - \frac{2^n}{\xi^n} \cdot (p_{\text{clean}}(g_\theta(x)|x) + L / \xi))^2}{1 +  \frac{2^n}{\xi^n} \cdot (p_{\text{clean}}(g_\theta(x)|x) + L / \xi)}  + \frac{1}{2} - \frac{2^{n-1}}{\xi^n} (p_{\text{clean}}(g_\theta(x)|x) + L / \xi),
\end{align*}
and upper-bounded by

\begin{align*}
\frac{1}{2} \cdot \frac{(1 - \frac{2^n}{\xi^n} \cdot (p_{\text{clean}}(g_\theta(x)|x) - L / \xi))^2}{1 +  \frac{2^n}{\xi^n} \cdot (p_{\text{clean}}(g_\theta(x)|x) - L / \xi)}  + \frac{1}{2} - \frac{2^{n-1}}{\xi^n} (p_{\text{clean}}(g_\theta(x)|x) - L / \xi).
\end{align*}

After taking Taylor expansion up to the linear term of \( 1/\xi^n \), we get that
\begin{align*}
1 - \frac{2^{n+1}}{\xi^n} p_{\text{clean}}(g_\theta(x)|x) + o(1/\xi^n)
\leq
    \chi^2_{\text{Pearson}} ( p_g^\xi  (p_{\text{clean}} + p_g^\xi)/ 2 ),
\end{align*}
and
\begin{align*}
    \chi^2_{\text{Pearson}} ( p_g^\xi  (p_{\text{clean}} + p_g^\xi)/ 2 ) 
\leq
1 - \frac{2^{n+1}}{\xi^n} p_{\text{clean}}(g_\theta(x)|x) + o(1/\xi^n).
\end{align*}

Thus,
\begin{align*}
    \chi^2_{\text{Pearson}} ( p_g^\xi || (p_{\text{clean}} + p_g^\xi) / 2 )  = 1 - \frac{2^{n+1}}{\xi^n} p_{\text{clean}}(g_\theta(x)|x) + o(1/\xi^n).
\end{align*}

As \( \xi \) is approaching \( +\infty \), the main term depending on the \( g_\theta(x) \) is \( - \frac{2^{n+1}}{\xi^n} p_{\text{clean}}(g_\theta(x)|x) \) which is minimized then \( g_\theta(x) = \underset{y}{\mathrm{arg}\,\max}\ p_{\text{clean}}(y|x) \) as global maximum is unique.
\end{proof}

\newpage
\section{Additional results for perceptual losses}

\subsection{More on motivation}
\label{app:motiv}
\begin{figure}[h!]
    \centering
    \begin{minipage}[t]{0.49\textwidth}
        \centering
        \includegraphics[height=4.5cm]{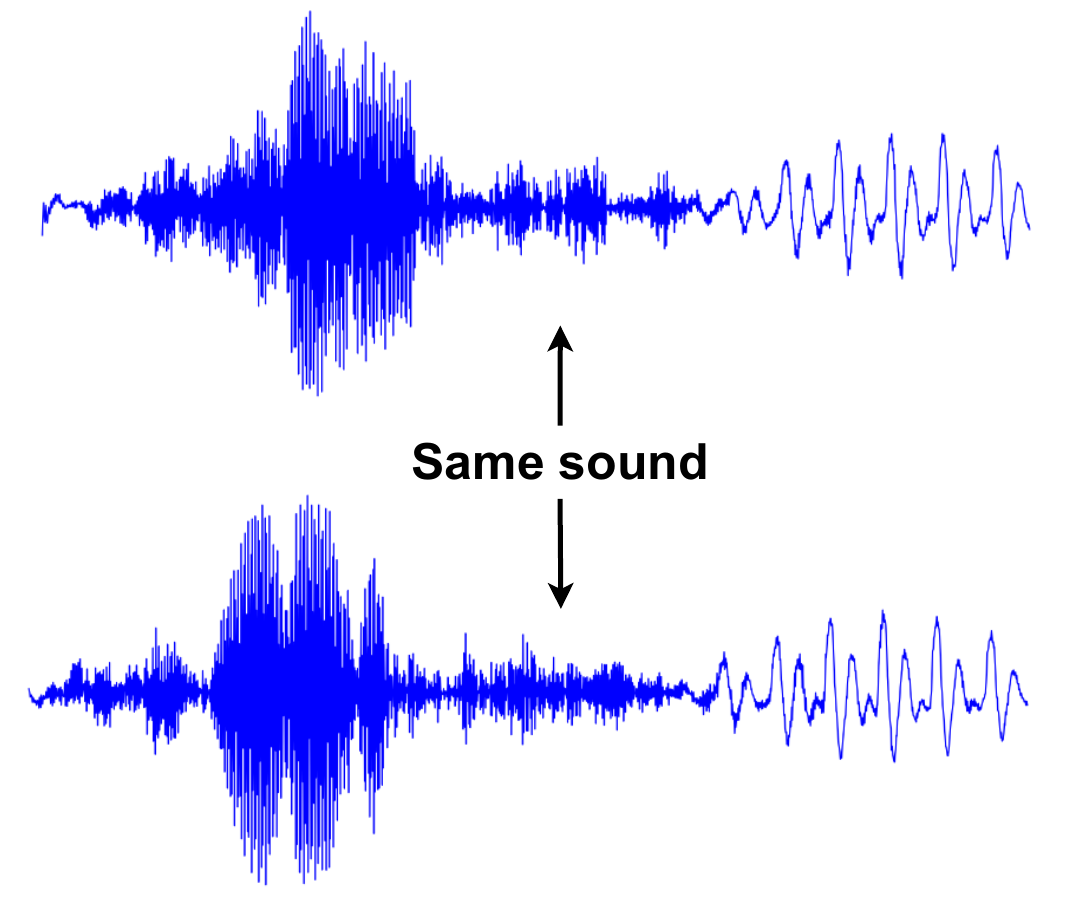}
        \caption{Ground truth waveform and waveform resynthesized by HiFi GAN vocoder. While waveforms significantly differ, they correspond to the same sound, creating ambiguity in generation.}
        \label{fig:waveforms}
    \end{minipage}
    \hfill
    \begin{minipage}[t]{0.49\textwidth}
        \centering
        \includegraphics[height=4.3cm]{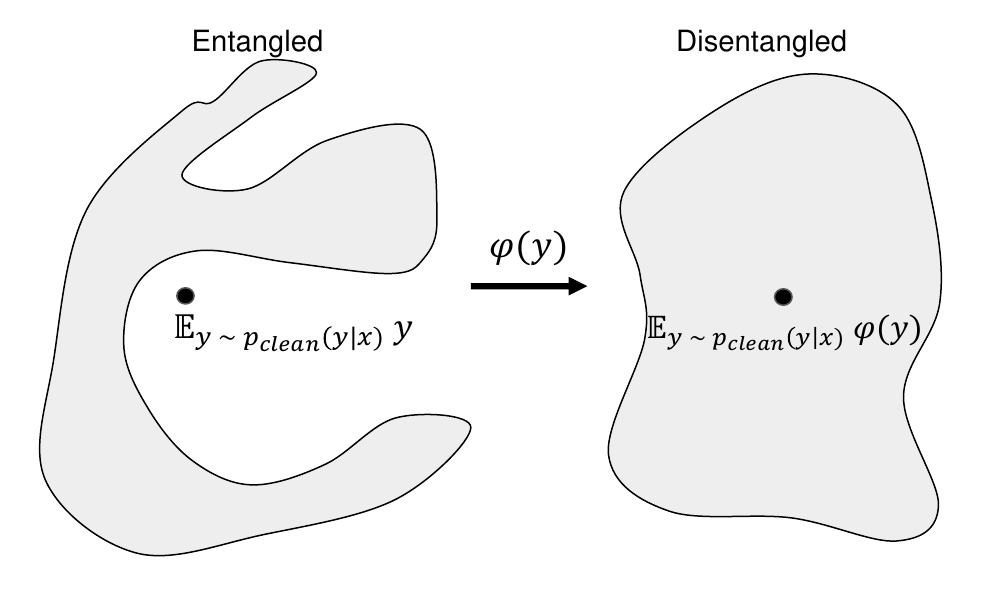}
        \caption{Regression in an entangled feature space might cause the expectation to lie outside the regions of high density, while regression in a disentangled space facilitates the expectation to lie within the regions of high probability density.}
        \label{fig:math_ill}
    \end{minipage}
\end{figure}

From the probabilistic point of view, minimization of the point-wise distance leads to an averaging effect. For example, optimization of the mean squared error between waveforms delivers the expectation of the waveform over the conditional distribution of clean speech given its degraded version \(g_\theta(x) = \mathbb{E}_{y \sim p_{\text{clean}}(y|x)}[y]\). 
The key thing is that the expectation over the distribution is not guaranteed to lie in the regions of high density of this distribution (see \cref{fig:waveforms} and \cref{fig:math_ill}).

Mathematically speaking, let \(\phi(x)\) be the mapping to this space, and assume there exists a reverse mapping \(\phi^{-1}(z)\) such that \(\phi^{-1}(\phi(x)) = x\). 
We would like to note that in practice, for regression purposes, there is no need to explicitly know the reverse mapping \(\phi^{-1}(z)\) as long as \(\phi(x)\) is differentiable.
The regression in the space produced by the mapping leads to averaging in this space; thus, after minimizing the MSE loss 
\[
\mathbb{E}_{x, y \sim p_{\text{clean}}(y) p(x|y)}\|\phi(y) - \phi(g_\theta(x))\|^2,
\]
one would obtain the solution 
\[
g_\theta(x) = \phi^{-1}\left(\mathbb{E}_{y \sim p_{\text{clean}}(y|x)}[\phi(y)]\right).
\]
Therefore, a desirable property for the \(\phi(x)\) mapping to be used as the regression space is that 
\[
\phi^{-1}\left(\mathbb{E}_{y \sim p_{\text{clean}}(y|x)}[\phi(y)]\right) = \underset{y}{\mathrm{arg\,max}}\, p_{\text{clean}}(y|x),
\]
i.e., we would like averaging in \(\phi\)-space to provide a representation corresponding to the most likely solution (the most probable reconstruction as discussed in the previous section).
In practice, this property is difficult to verify. 
Moreover, finding such a mapping is likely to be a task which is not easier than the original problem.
Therefore, based on this intuition, we propose some heuristic rules for assessing the structure of regression spaces produced by different mappings.

\label{app:spec_loss}
\begin{figure}[!ht]
        \centering
        \includegraphics[height=19.5cm]{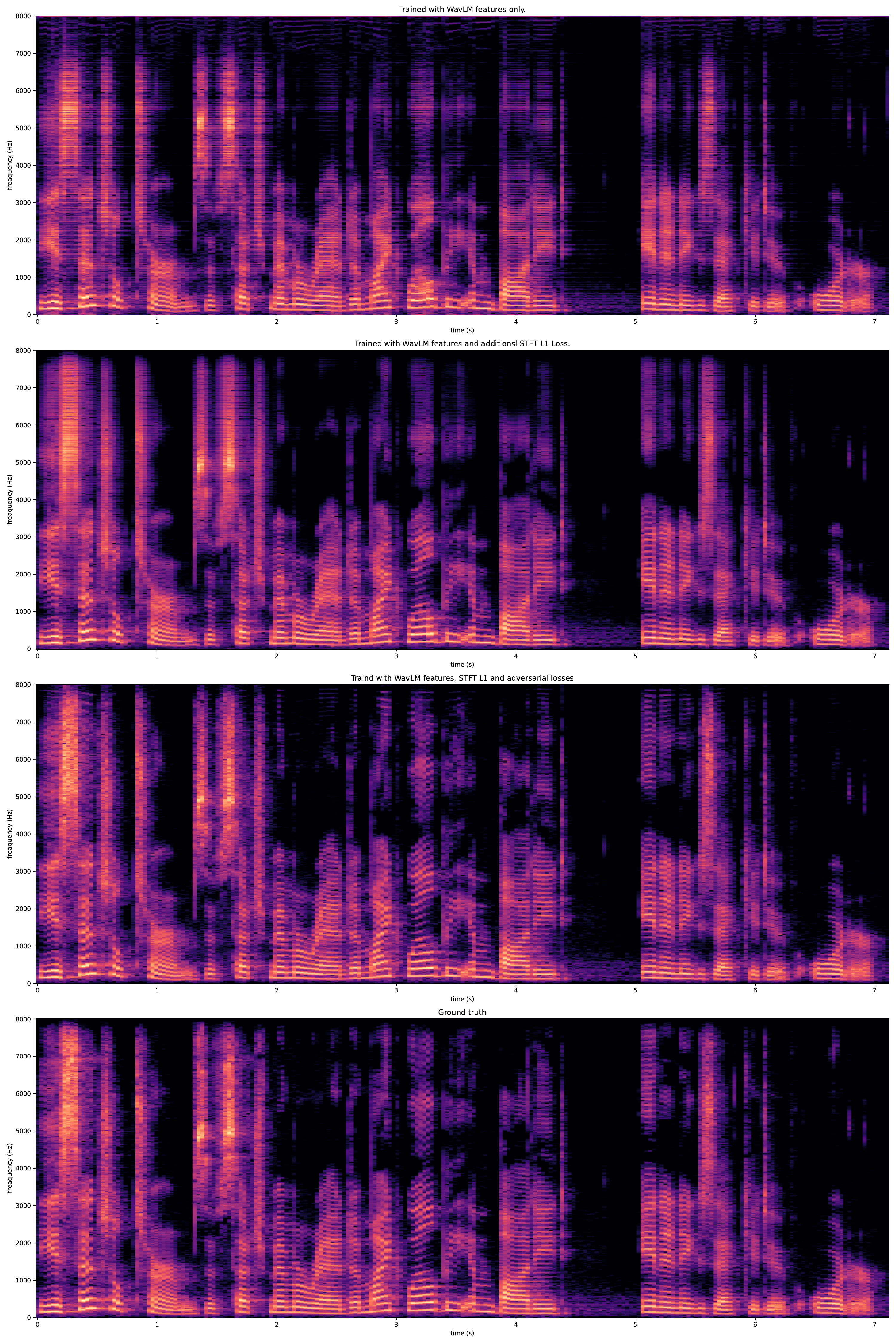}
        \caption{Comparison of training schemes on spectrograms. Going from the top to the bottom, 
        the first spectrogram is obtained from the model trained only on WavLM \citep{chen2022wavlm}
        features, the second one is produced by the model trained on both WavLM features and STFT L1 loss. The third spectrogram is obtained by training with both mentioned losses and adversarial loss. The last spectrogram is computed with ground truth audio.
        }
        \label{fig:specs}
\end{figure}

\subsection{Influence of losses on final audio}
To illustrate the influence of LMOS loss, as well as the importance of adversarial training, we examine the impact of different training setup using vocoding and discuss the results using spectrograms in \cref{fig:specs}. We start with training the vocoding model using only convolutional features from  WavLM. This training turns out to be suboptimal, as the model produces noticeable artefacts, visible in the top spectrogram in \cref{fig:specs}. Next, we complement the WavLM feature loss with the STFT L1 loss, which is a L1 distance between spectrograms of reference and predicted audios. This method yields better quality, however some artefacts still remain. Notice, that such training particularly struggles to reconstruct high frequencies and capture the harmonic content. Finally, we train the model using adversarial loss. For this experiment we use original HiFi GAN \citep{kong2020hifi} setup. Adversarial training helps to alleviate artefacts and produces better quality, as compared to the regression training used in both former experiments. For these reasons, we rely both on our newly developed reconstruction loss and adversarial training for creating our speech enhancement model. 

\subsection{Comparison with prior work}
\label{app:vocod}
We additionally compare the LMOS against other speech perceptual losses from the literature: \citep{hsieh2020improving, close2023effect, defossez2020real, kong2020hifi}. The comparison is outlined in \cref{table:mos_scores_voc}. The experiments are conducted on neural vocoding task using LJSpeech dataset \citep{ljspeech17}. The HiFi-GAN generator V1 was trained for 1M iterations with batch size 16 for each loss function.

\begin{table}[h]
    \centering
        \caption{MOS on neural vocoding for different loss functions.}
    \begin{tabular}{lc}
        \toprule
        \textbf{Method} & \textbf{MOS ($\uparrow$)} \\ 
    \midrule
    \midrule
       PFPL \citep{hsieh2021improving} & 2.40 $\pm$ 0.08 \\ 
        SSSR loss with HuBERT features \citep{close2023effect} & 2.58 $\pm$ 0.08 \\
        MS-STFT + L1 waveform \citep{defossez2020real} & 3.16 $\pm$ 0.08 \\ 
        LMOS (ours) & 4.21 $\pm$ 0.07 \\ 
        adv. MPD-MSD \citep{kong2020hifi} & \textbf{4.65 $\pm$ 0.04} \\ 
        Ground Truth & 4.66 $\pm$ 0.04 \\ \bottomrule
    \end{tabular}

    \label{table:mos_scores_voc}
\end{table}

\subsection{WavLM layers}
\label{app:wavlm_l}

To find suitable layer of WavLM \citep{chen2022wavlm} model for our loss function, we take activations of each layer and measure Rand index and correlation with SNR. We seek to find such layer, that would produce high score for both metrics. Intuitively, this suggests that the layer's activations create a space where audio with varying acoustics is effectively separated, and their hidden representations are responsive to noise.

\begin{figure}[h!]
    \centering
    \includegraphics[height=4.5cm]{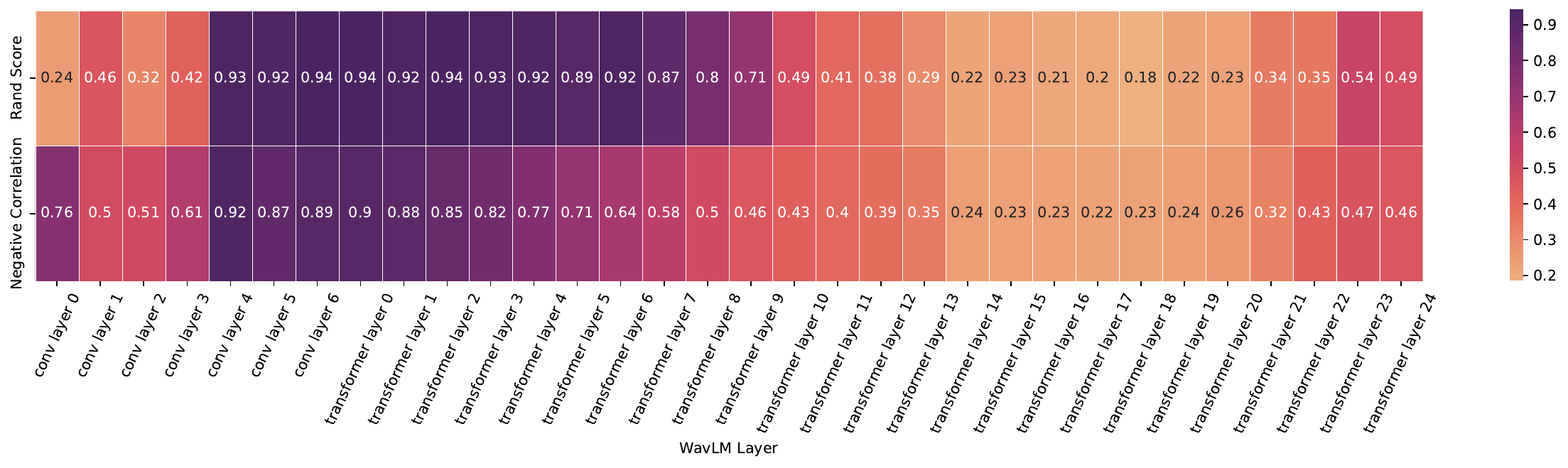}
    \caption{Comparison of WavLM features using Rand index and negative correlation of SNR}
    \label{fig:wavlm_layers}
\end{figure}

As we see in \cref{fig:wavlm_layers}, the most suitable features come from the convolutional encoder or first transformer layer. For convenience, we compute loss using features from convolutional encoder (we call the these outputs WavLM-Conv features).

\subsection{WavLM features visualization}
\begin{figure}[!h]
    \centering
        \centering
        \includegraphics[height=6cm]{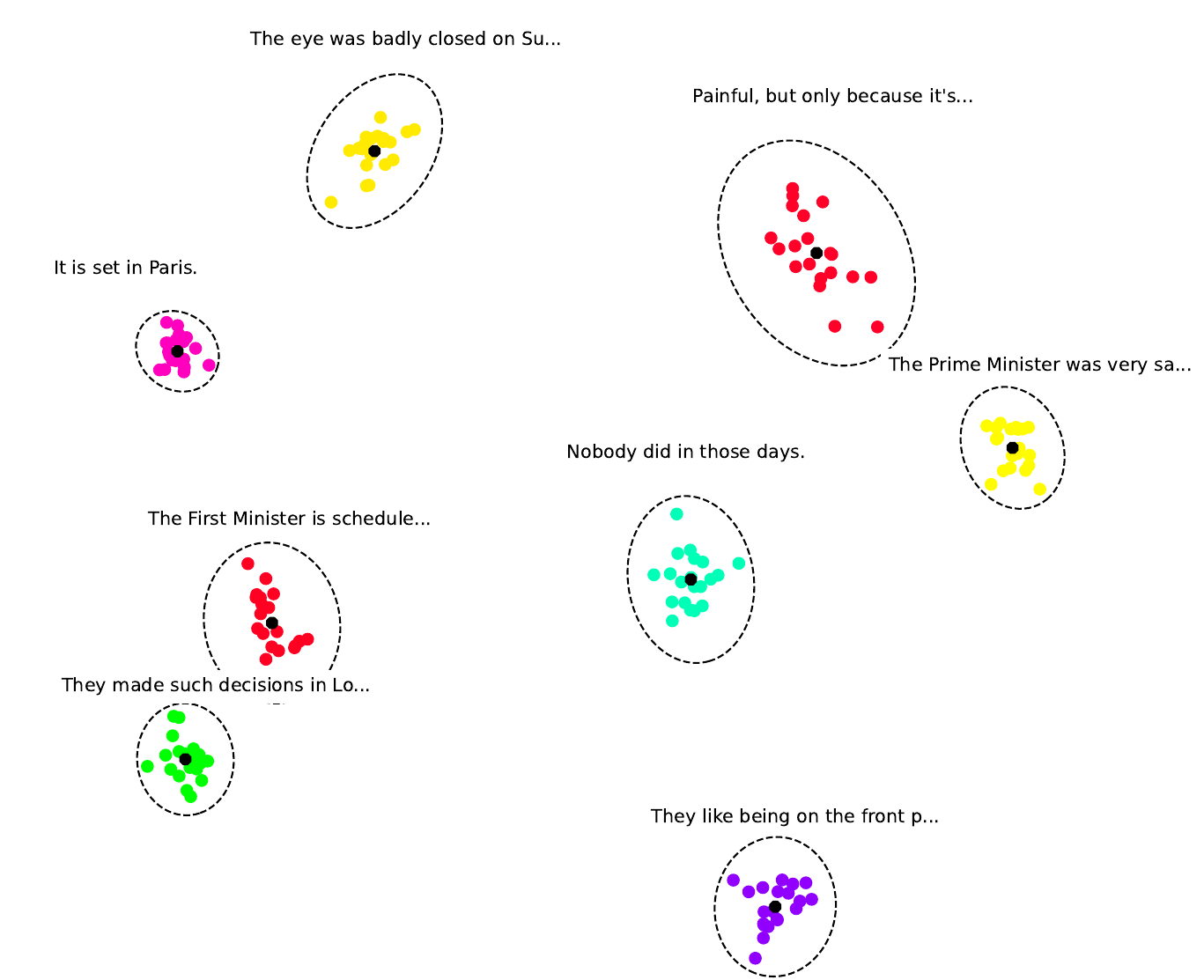}
        \caption{Clustering rule visualized on Wavlm-Conv PCA features. The phrases corresponding to each cluster are visualized.}
        \label{fig:clust_real}
\end{figure}

\begin{figure}[!h]
        \centering
        \includegraphics[height=6cm]{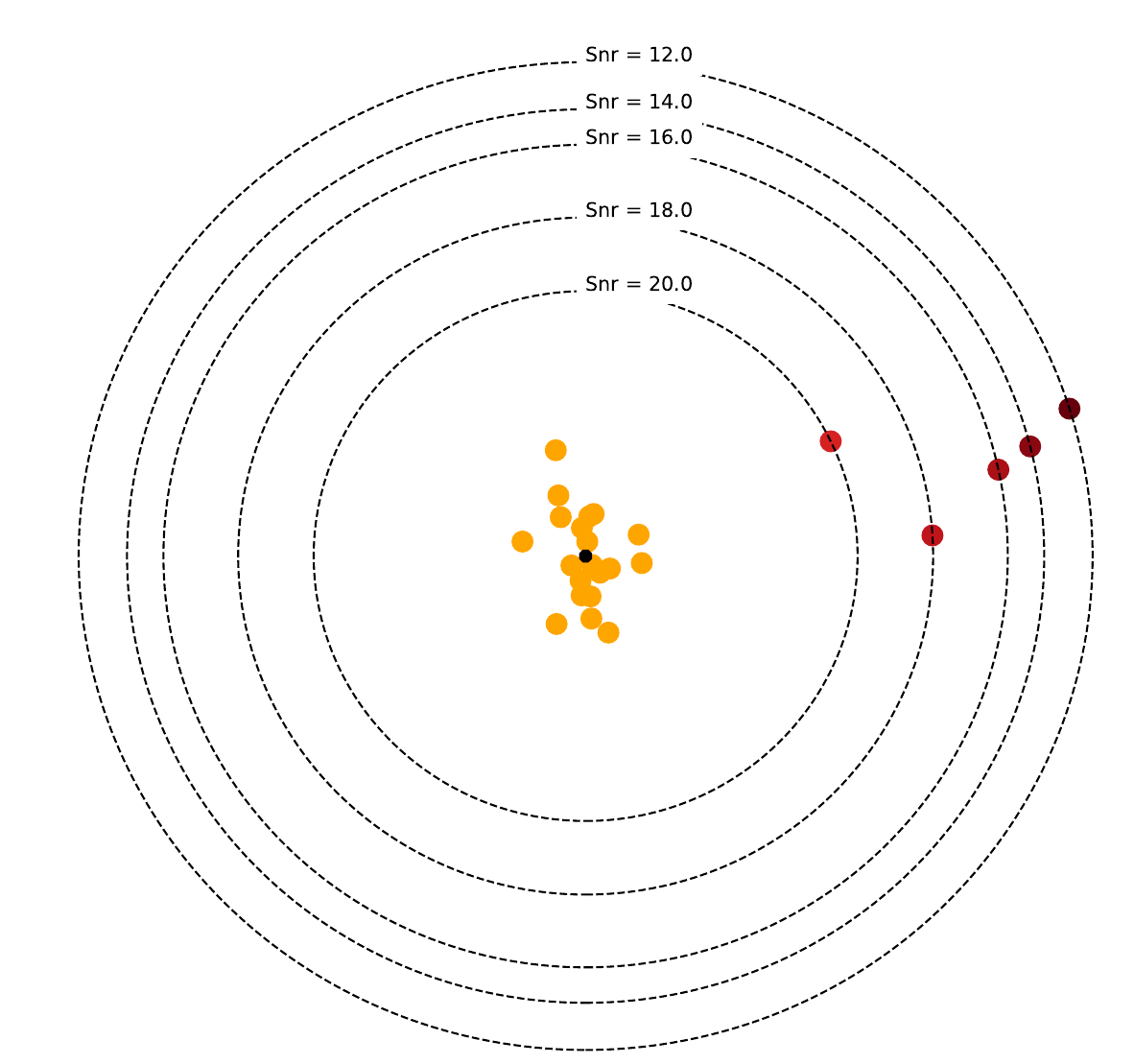}
        \caption{SNR rule visualized for Wavlm-Conv PCA features for a particular cluster.}
        \label{fig:snr_real}
\end{figure}

To provide visual evidence for the successful choice of WavLM features, as well as to illustrate \textbf{clustering rule} and \textbf{SNR rule}, we show the first two principal components of WavLM-Conv features and depict corresponding clusters (\cref{fig:clust_real}) and noisy features (\cref{fig:snr_real}). 
\Cref{fig:clust_real} illustrates that audio samples with the same lexical content form well-defined clusters, whereas those with different content are properly disentangled. \Cref{fig:snr_real} shows that increasing SNR in audio moves it away for the centre of the cluster. These figures provide empirical evidence for the heuristics discussed in \cref{section:loss_criteria}

\newpage
\section{Comparison with MIIPHER}
\label{app:miip}
We compare the enhancement quality of LibriTTS test\_other samples as released by the Miipher authors~\citep{koizumi2023libritts}. The results are outlined in \cref{table:mos_scores_m}. For the information on WER refer to \cref{app:metric_computations}. computations.

\begin{table}[h]
    \centering
        \caption{MOS Scores for comparison with Miipher (LibriTTS, test other).}
    \label{table:mos_scores_m}
    \begin{tabular}{l c c c c c c}
        \toprule
        Method & MOS ($\uparrow$) & UTMOS ($\uparrow$) & DNSMOS ($\uparrow$) & WV-MOS ($\uparrow$) & WER ($\downarrow$) \\
        \midrule
        Input & 3.59 $\pm$ 0.07   & 3.41 $\pm$  0.13 & 2.88 $\pm$  0.12 & 3.69 $\pm$  0.11 & 0.23 \(\pm\) 0.04 \\
        
        Miipher & 4.18 $\pm$ 0.06  & 3.95 $\pm$  0.13 &  2.99 $\pm$  0.12 & 4.15 $\pm$  0.14 & 0.22 \(\pm\) 0.05 \\
        
        Ours (16 kHz) & 4.24 $\pm$ 0.05  & \textbf{4.18 $\pm$  0.08} & \textbf{3.15 $\pm$ 0.09} & \textbf{4.28 $\pm$  0.10} & \textbf{0.18 \(\pm\) 0.04}  \\
        
        Ours (48 kHz) & \textbf{4.54 $\pm$ 0.05}  & \textbf{4.18 $\pm$  0.08} & \textbf{3.15 $\pm$  0.09 } & \textbf{4.28 $\pm$  0.10} & \textbf{0.18 \(\pm\) 0.04}  \\
        \bottomrule
    \end{tabular}

\end{table}
\newpage
\section{Implementation details}

\subsection{Reproducibility statement}
\label{app:repr_statement}

Our model is based on HiFi++ architecture \citep{andreev2022hifi++}, for which code is available in \url{https://github.com/SamsungLabs/hifi_plusplus}. We used several open-source datasets for training. More specifically, we used
\href{https://ccrma.stanford.edu/~gautham/Site/daps.html}{DAPS},
\href{https://www.openslr.org/60}{LibriTTS-R} and 
\href{https://github.com/microsoft/DNS-Challenge}{Deep Noise Suppression Challenge (DNS)} (only noises).
We rely on official implementations of MS-STFT discriminators, which can be found in \url{https://github.com/facebookresearch/encodec}. For LMOS loss and HiFi++ conditioning, we used WavLM large  model from Hugging Face, which can be found in \url{https://huggingface.co/microsoft/wavlm-large}.

\subsection{Augmentations}

Before mixing with noise, we convolve the speech signal with a randomly chosen microphone impulse response from the Microphone Impulse Response project\footnote{\url{http://micirp.blogspot.com/}} and apply other digital distortions. 
With a probability of 0.8, we also convolve the signal with a room impulse response randomly chosen from those provided in the DNS dataset. We additionally apply audio effects from the torchaudio library~\citep{hwang2023torchaudio}, trying to simulate digital distortions. Parameter values are chosen randomly; only one codec is applied. 
\label{app:augs}
\begin{table}[!h]
    \centering
    \caption{Applied augmentations.}
  \label{tab:phase}
  \begin{tabular}{l c c c }
    \toprule
    Augmentation & Prob. &  Param. name & Interval of values  \\
    \midrule
    \midrule
             acrusher &  0.25 & bits & [1,9]  \\ 
             crystalizer &  0.4 & intensity	& [1,4]  \\
             flanger & 0.15  & depth	& [1,8]  \\
            vibrato & 0.15 & frequency	& [5,8]  \\
            \midrule
           \multirow{2}{*}{\shortstack[l]{codec ogg \\ codec mp3}} &  \multirow{2}{*}{ 0.45}  & encoder	& vorbis, opus  \\ & & bit rate  & [4000,16000] \\
      \bottomrule
       
  \end{tabular} 
   
  \label{table:augs}
\end{table}

\subsection{Model architecture}
In this section we detail parameters for our speech enhancement models. We use HiFi++ \citep{andreev2022hifi++} as backbone. This architecture consists of four parts: Spectral UNet, HiFi Upsampler, WaveUNet and SpectalMaskNet. In addition to that, we use WavLM-large feature encoder with transformer features, which takes a waveform as an input and outputs the last hidden state of the transformer, which is stacked with the SpectralUNet output and fed into the HiFi Upsampler. In the next paragraphs, we thoroughly discuss the architecture of each component. Architecture summary is presented in \cref{table:model_architecture}

\paragraph{Residual blocks}
To better understand Spectral UNet, WaveUNet and SpectralMaskNet we first discuss the key building block. 
Each residual block is composed of convolution, followed by LeakyReLU. We use weight norm \citep{weightnorm} instead of BatchNorm, we found it to be more efficient in our preliminary experiments. We use kernel size 3 for 2D UNets and kernel size 5 for 1D UNet. Each residual block has additive skip connection. In our architecture, we use a stacked composition of residual blocks. The number of residual blocks stacked together in one layer is called depth. 

\paragraph{SpectralUNet architecture}
SpectralUNet processes a mel-spectrogram input with dimensions [B, 80, T] and outputs a hidden state with dimensions [B, 512, T]. The input mel-spectrogram is combined with positional encoding. The architecture is based on a 2D UNet \citep{RonbergerUNet} with five layers, each having respective channels of $[16, 32, 64, 128, 256]$ and a depth of 4. Additional convolutions are applied after these layers to transform 2D data from shape [B, 16, W, T] to [B, 1, W, T], and then to [B, 512, T]. This output is concatenated along the channel dimension with the final transformer hidden state of WavLM \citep{chen2022wavlm}, which has been interpolated using the nearest-neighbour method to match the output length of SpectralUNet in the time dimension. The concatenated result then passes through a Residual block with kernel size 3 and with a width of 1536 (1024 from the final transformer hidden state and 512 from SpectralUNet output), then passes through 1d Convolution with kernel size 1 and LeakyReLU to obtain size 512 in the channel dimension, and is subsequently fed into the HiFi Upsampler.

\paragraph{HiFi Upsampler architecture}

For HiFi Upsampler we use HiFi generator architecture from \citep{kong2020hifi}. In our implementation
we use 4 layer model with upsample rates $[8, 8, 2, 2]$, kernel size $[16, 16, 4, 4]$, hidden sizes $[512, 256, 128, 64]$.
For each layer we use residual blocks with kernels $[3, 7, 11]$ and dilations $[(1, 3, 5), (1, 3, 5), (1, 3, 5)]$.

\paragraph{WaveUNet architecture}
We implement WaveUNet following \citep{waveUNet}.  We use 4-layer WaveUNet with channels $[128, 128, 256, 512]$, each layer has depth 4. Our WaveUNet takes the concatenation of input waveform and HiFi Upsampler as input. In this way we provide a residual connection for Spectral UNet and HiFi Upsampler. 

\paragraph{SpectralMaskNet architecture}
The final Stage of our pipeline is SpectralMaskNet. It applies channel-wise Short Time Fourier Transform (STFT) to the input, decomposes it into phase and amplitude, and then processes amplitude with 2D UNet. Finally, it multiplies processed amplitude and phase and applies inverse STFT to obtain the final audio. For processing amplitude 2D UNet with channels $[64, 128, 256, 512]$ as a backbone, each layer has depth 1. The output of SpectralMaskNet is the final output of our 16kHz model.

\paragraph{WaveUNet Upsampler}
In order to enable the model to output samples in 48kHz we also add WaveUNet Upsampler, that upsamples the final audio. For that task we use WaveUNet with 5 layers and channels $[128, 128, 128, 128, 256]$, each layer has depth 3. The outputs of each layer are downsampled with the scale 4. After the output of the final upsample layer, we have an additional head that has 512 features and is used for directly upscaling the WaveUNet output into 48kHz.

\paragraph{Architecture summary}
More details about the architecture can be found in \citep{andreev2022hifi++}. We present the summary of parameters for Spectral UNet, WaveUNet and SpectralMaskNet  in \cref{table:model_architecture}.

\begin{table}[!h]
    \caption{Summary of parameters.}
    \label{table:model_architecture}
    
    \centering
    \begin{tabular}{lcccc}
        \toprule
        
        \mlcell{Submodel} & \mlcell{Channel Sizes}  & \mlcell{Kernel Size \\ (constant for \\ each layer)}  & \mlcell{Layer Depth \\ (constant for \\ each layer)} \\
        \midrule
        
        \midrule
        Spectral UNet              & \([16, 32, 64, 128, 256]\) & \(3\) & \(4\)  \\
        HiFi Upsampler             & \([512, 256, 128, 64]\) & \([3,7,11]\) & \(3\)  \\
        WaveUNet                   & \([128, 128, 256, 512]\) & \(5\) & \(4\) \\
        Spectral MaskNet          & \([64, 128, 256, 512]\) & \(3\) & \(1\) \\
        WaveUNet Upsampler     & \([128, 128, 128, 128, 256]\) + 512 & \(5\) & \(3\)  \\
        \bottomrule
    \end{tabular}
\end{table}

\paragraph{Discriminators}

We employ MS-STFT discriminators \citep{defossez2023high} for adversarial training. We use 5 MS-STFT discriminators with n\_ffts $[ 4096, 2048, 1024, 512, 256 ]$, 
hop\_lengths $[ 1024, 512, 256, 128, 64]$, win\_lengths $[ 4096, 2048, 1024, 512, 256 ]$ in 48kHz setting, for 16kHz all above-mentioned parameters are divided by 2. Our implementation closely follows \citep{defossez2023high}.

\paragraph{Training details} The main model is trained using 8 Nvidia P40 GPUs with effective batch size 32 and AdamW \citep{AdamW} as our main optimizer. For pretraining we use learning rate 0.0002, betas (0.8, 0.99) and learning rate exponential decay of 0.996 for each 200 iterations, the pretraining lasts 100,000 iterations. For the second stage we use learning rate 0.0002 with betas (0.8, 0.99) and learning rate decay 0.995, whereas for discriminators we use learning rate 0.0002 with betas (0.5, 0.999) and learning rate decay 0.995, the discriminators perform 2 optimization iterations for every 1 generator's optimization iteration, the training lasts 30,000 generator's iterations. We also use a linear warm-up for 2000 iterations for the generator. We use the same training parameters for the third stage during generator's 40,000 iterations. The first and the second stage are trained in 16kHz using LibriTTS-R dataset with noises from DNS and augmentations, discussed in \cref{app:augs}. Whereas the third stage is trained in 48kHz using DAPS dataset with noises from DNS as well. Datasets are discussed in \cref{app:repr_statement}

\paragraph{
    Comparison of the datasets and resources used to train FINALLY and baseline models.
}
To better contextualize our results, we provide a detailed comparison of the model sizes and training data of the baseline models. 

\begin{table}[h!]
\centering
\label{tab:resource_comparison}
\caption{
    Comparison of resources and data used for training.
    }
\scalebox{0.80}{
\begin{tabular}{lccc}
\toprule
Model       & Training Data Scale (clean data) & Model Size (parameters) & RTF on V100 GPU \\
\midrule
VoiceFixer  & 44 hours (VCTK)                  & 112 M                   & 0.02            \\
DEMUCS      & 500 hours (DNS)                  & 61 M                    & 0.08            \\
STORM       & 200 hours (WSJ0 and VCTK)        & 28 M                    & 1.05            \\
BBED        & 140 hours (WSJ0)                 & 65 M                    & 0.43            \\
HIFI-GAN-2  & 5 hours (DAPS)                   & 34 M                    & 0.50            \\
Universe    & 1500 hours (private data)        & 189 M                   & 0.50            \\
FINALLY (ours) & 200 hours (LibriTTS-R and DAPS) & 454 M (including 358 M of WavLM) & 0.03 \\
\bottomrule
\end{tabular}}
\end{table}

Although our model has more parameters compared to the baseline models, the majority of these parameters are dedicated to handling low-resolution features. For instance, the Transformer in WavLM processes representations of the waveform that are downsampled by a factor of 320, operating at 50 Hz. On the other hand, models such as HiFi-GAN-2 primarily work at the full waveform resolution, similar to WaveNet. This design enables our model to be more efficient in terms of computational resources, resulting in a significantly lower Real-Time Factor (RTF).

\subsection{Limitations}
The primary area for further developing in the proposed model lies in improving perceptual quality, particularly in instances when the speech is severely degraded. Although our model is capable of capturing the content of such signals, even in situations that may be impossible for humans, some artefacts remain. These artefacts could possibly be attributed to the high uncertainty of the initial signal, such as the speaker's voice, for instance.

Another crucial aspect to consider is the streaming mode. While the proposed model is fast, it doesn't suit low latency scenarios, which are widely applicable in telecommunications. The big obstacle to such improvement could be the WavLM \citep{chen2022wavlm} model that utilizes context for identifying the content of speech. This model has the boundaries of the required context and, consequently, the minimum latency achievable.

\section{Evaluation details}
\label{app:eval}

\subsection{Metric computation}
\label{app:metric_computations}

To compare models in their ability to correctly reconstruct the speech, we use Phoneme Error Rate (PhER) and Word Error Rate (WER). Given the reference audio and audio, produced by the speech enhancement model, metrics are computed in the following way. First, Automatic Speech Recognition (ASR) model is applied to both audios, yielding phoneme representations. Then, the editing distance is computed over predicted phonemes (or words for WER) for the reference and the predicted audio. The distance is normalized to the length of the reference phoneme representation (or the length of the sentence for WER) and averaged over the validation set. We rely on \url{https://huggingface.co/} for the ASR models and metric computations. For PhER we use \href{https://huggingface.co/mrrubino/wav2vec2-large-xlsr-53-l2-arctic-phoneme}{mrrubino/wav2vec2-large-xlsr-53-l2-arctic-phoneme} model, for WER we use \href{https://huggingface.co/jonatasgrosman/wav2vec2-large-xlsr-53-english}{jonatasgrosman/wav2vec2-large-xlsr-53-english}. These are Wav2Vec2-based \citep{baevski2020wav2vec} models for ASR.

\subsection{Ablation details}
\label{app:ablation_details}

For the ablation study, we utilized a smaller version of our original model to expedite evaluations, with the reduced model containing approximately 22 million parameters. The dataset used for this ablation was provided by \citep{nagrani2017voxceleb}. The ablation process was carried out in several stages. Initially, we compared the performance of our LMOS loss with L1Spec loss \citep{kong2020hifi}, MS-STFT loss \citep{defossez2020real}, and RecLoss \citep{defossez2023high} during the pretraining phase. Each loss function was used independently to train the model in a regressive manner for 500,000 steps with a batch size of 12, utilizing two Nvidia P40 GPUs. Our findings indicated that LMOS consistently outperformed the other losses. Detailed results are presented in \cref{tab:comparison_metrics_stage1}. It's important to note that MOS evaluations were not conducted during the first stage, as the enhanced audio typically displayed artefacts following this stage of training.

\begin{table}[h]
    \centering
    \label{tab:comparison_metrics_stage1}
    \caption{Ablation for regression losses.}
    \begin{tabular}{lccc}
        \toprule
        Metric   & UTMOS ($\uparrow$) & WV-MOS ($\uparrow$) & DNSMOS ($\uparrow$) \\
        \midrule
        MS-STFT  & \(2.54 \pm 0.10\) & \(2.77 \pm 0.10\) & \(3.04 \pm 0.05\) \\
        RecLoss  & \(2.53 \pm 0.10\) & \(2.77 \pm 0.10\) & \(3.04 \pm 0.05\) \\     
        L1Spec   & did not converge  & did not converge  & did not converge \\
        LMOS     & \(\mathbf{3.43 \pm 0.09}\) & \(\mathbf{3.57 \pm 0.05}\) & \(\mathbf{3.14 \pm 0.04}\) \\
        \bottomrule
    \end{tabular}
\end{table}

Next, we ablated the use of mentioned losses in purely adversarial setup without pretraining. The same small model was trained from scratch using one of the discussed reconstruction losses with adversarial LS-GAN loss \citep{mao2017least} and feature matching loss \citep{kong2020hifi}. We used MS-STFT discriminators \citep{defossez2023high} for adversarial training.  We fixed coefficients for the feature matching loss at 5 and for the LS-GAN loss at 15, and then grid searched the coefficient for each reconstruction loss. We experimented with the possible coefficient to be 1, 3, 10, 20, 50 and found that is all cases the best coefficient is 3. We found that LMOS loss outperforms other reconstruction losses in this setup as well. 

In the first stage we experimented with the addition of SSL features to boost model quality. For these experiments we took our best model trained with LMOS loss. As our ablation indicates, transformer features of WavLM significantly increase the model quality. 

Finally, we train our big model, using the ablated architecture. We report the results in 'Scaling' row in \cref{table:loss_ablations}. We also show how the introduction of the WaveUNet Upsampler and Human Feedback (HF) losses influence the final quality of our model.

In conclusion, we found that the best model is the one, trained with WavLM transformer features  and LMOS reconstruction loss. Moreover, we found, that using HF losses also significantly increases the final quality.

\paragraph{Comparison with HiFi++.}

Since our model is based on HiFi++ architecture \citep{andreev2022hifi++}, use explicitly show how the improvements we introduced increase the  quality of enhanced audio compared to HiFi++ baseline using VoxCelex dataset ~\citep{nagrani2017voxceleb}. The results are presented in ~\cref{tab:comparison}

\begin{table}[ht]
\centering
\caption{Comparison with HiFi++ on VoxCeleb data.}
\label{tab:comparison}
\begin{tabular}{lccc}
\toprule
Model & UTMOS ($\uparrow$) & WV-MOS ($\uparrow$) & DNSMOS ($\uparrow$) \\
\midrule
Input          & $2.72 \pm 0.11$ & $2.90 \pm 0.16$ & $2.72 \pm 0.11$ \\
HiFi++         & $2.76 \pm 0.13$ & $2.68 \pm 0.14$ & $2.98 \pm 0.07$ \\
FINALLY (ours) & $\mathbf{4.05 \pm 0.07}$ & $\mathbf{3.98 \pm 0.06}$ & $\mathbf{3.31 \pm 0.04}$ \\
\bottomrule
\end{tabular}
\end{table}

\section{Subjective evaluation}

\begin{figure}[h]
\begin{center}
\includegraphics[width=0.9\linewidth]{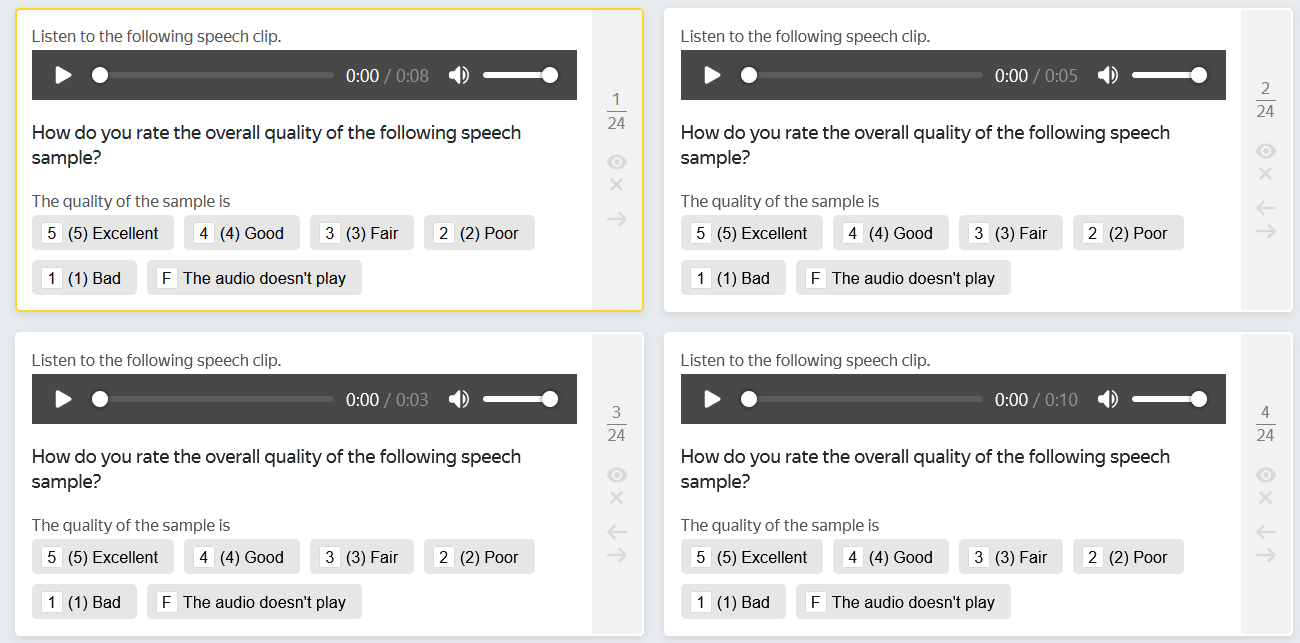}
\end{center}
\caption{The assessor's interface.}
\label{fig:instructions}
\end{figure}
\begin{figure}[h]
\begin{center}
\includegraphics[width=0.9\textwidth]{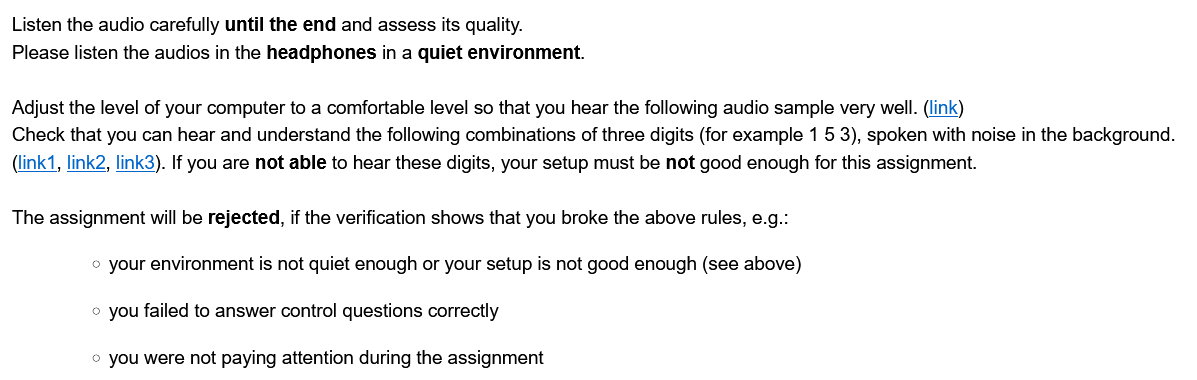}
\end{center}
\caption{The rules for the assessor.}
\label{fig:interface}
\end{figure}
\label{app:subj}
We measure mean opinion score (MOS) of the model using a crowdsourcing adaptation of the standard absolute category rating procedure.
Our MOS computing procedure is as follows.

\begin{enumerate}
    \item Select a subset of 40 random samples from the test set (once per problem, i.e. for bandwidth extension or speech enhancement).
    \item Select a set of models to be evaluated; inference their predictions on the selected subset.
    \item Randomly mix the predictions and split them into the pages of size 20 almost uniformly.
          Almost uniformly means that on each page there are at least $\lfloor \frac{20}{\mathrm{num\_models}} \rfloor$ samples from each model.
    \item Insert additional 4 trapping samples into random locations on each page: 2 samples from ground truth, and 2 samples of a noise without any speech.
    \item Upload the pages to the crowdsourcing platform, set the number of assessors for each page to at least 30.\\
          Assessors are asked to work in headphones in a quiet environment; they must listen to the audio until the end before assess it.
    \item Filter out the results where the ground truth samples are assessed with anything except 4 (good) and 5 (excellent), or the samples without voice are assessed with       anything except 1 (bad).
    \item Compute 95\% confidence intervals via bootstrapping.
\end{enumerate}

Since the models are distributed uniformly among the pages, assessor's biases affect all models in the same way, so the relative order of the models remains.
On the other hand, the assessor will have access to all variety of the models on one page and thus can scale his ratings better.
The other side is that the models' ratings depend on each other in this setting, because assessors tend to estimate the sample quality relatively to the average sample of the page. This means that when some models perform poorly in comparison, the higher scores are given to the better models.
4 trapping samples per page is also a reasonable choice, because one cannot just random guess the correct answers for these questions.

The instructions for the assessors and the screenshots of the evaluation interface are provided on \cref{fig:instructions} and, \cref{fig:interface} respectively.

The total amount of money spent on the surveys while preparing the paper is somewhere between \$1000 and \$1500 (it is hard to estimate exactly because there were a lot of exploration tests during creating the model and writing the paper). According to the crowdsourcing system statistics, the average hourly wage varies between tasks from \$2.5 to \$4, which exceeds by a large margin the minimal wage in the countries where the test was conducted.


\newpage
\newpage

\section*{NeurIPS Paper Checklist}


\begin{enumerate}

\item {\bf Claims}
    \item[] Question: Do the main claims made in the abstract and introduction accurately reflect the paper's contributions and scope?
    \item[] Answer: \answerYes{} 
    \item[] Justification: -

\item {\bf Limitations}
    \item[] Question: Does the paper discuss the limitations of the work performed by the authors?
    \item[] Answer:  \answerYes{} 
    \item[] Justification: See Appendix D.4

\item {\bf Theory Assumptions and Proofs}
    \item[] Question: For each theoretical result, does the paper provide the full set of assumptions and a complete (and correct) proof?
    \item[] Answer:\answerYes{} 
    \item[] Justification: Please see the Appendix A

    \item {\bf Experimental Result Reproducibility}
    \item[] Question: Does the paper fully disclose all the information needed to reproduce the main experimental results of the paper to the extent that it affects the main claims and/or conclusions of the paper (regardless of whether the code and data are provided or not)?
    \item[] Answer: \answerYes{} 
    \item[] Justification: Please see the Appendix C.

\item {\bf Open access to data and code}
    \item[] Question: Does the paper provide open access to the data and code, with sufficient instructions to faithfully reproduce the main experimental results, as described in supplemental material?
    \item[] Answer: \answerNo{} 
    \item[] Justification: We do not open the code for training of the model due to our organization policy, however, we use only publicly available data and most of the source codes are available as stated in Appendix D. Therefore, we expect that the work could be reproduced with moderate efforts.

\item {\bf Experimental Setting/Details}
    \item[] Question: Does the paper specify all the training and test details (e.g., data splits, hyperparameters, how they were chosen, type of optimizer, etc.) necessary to understand the results?
    \item[] Answer: \answerYes{} 
    \item[] Justification: Please see the Appendix D

\item {\bf Experiment Statistical Significance}
    \item[] Question: Does the paper report error bars suitably and correctly defined or other appropriate information about the statistical significance of the experiments?
    \item[] Answer: \answerYes{}  
    \item[] Justification: We report the confidence intervals computed by bootstrapping.

\item {\bf Experiments Compute Resources}
    \item[] Question: For each experiment, does the paper provide sufficient information on the computer resources (type of compute workers, memory, time of execution) needed to reproduce the experiments?
    \item[] Answer: \answerYes{} 
    \item[] Justification: please see the Appendix D
    
\item {\bf Code Of Ethics}
    \item[] Question: Does the research conducted in the paper conform, in every respect, with the NeurIPS Code of Ethics \url{https://neurips.cc/public/EthicsGuidelines}?
    \item[] Answer: \answerYes{} 
    \item[] Justification: -

\item {\bf Broader Impacts}
    \item[] Question: Does the paper discuss both potential positive societal impacts and negative societal impacts of the work performed?
    \item[] Answer: \answerNA{} 
    \item[] Justification: -
    
\item {\bf Safeguards}
    \item[] Question: Does the paper describe safeguards that have been put in place for responsible release of data or models that have a high risk for misuse (e.g., pretrained language models, image generators, or scraped datasets)?
    \item[] Answer:  \answerNA{} 
    \item[] Justification: -

\item {\bf Licenses for existing assets}
    \item[] Question: Are the creators or original owners of assets (e.g., code, data, models), used in the paper, properly credited and are the license and terms of use explicitly mentioned and properly respected?
    \item[] Answer: \answerYes{} 
    \item[] Justification: Please the Section 5.

\item {\bf New Assets}
    \item[] Question: Are new assets introduced in the paper well documented and is the documentation provided alongside the assets?
    \item[] Answer: \answerNA{} 
    \item[] Justification: -

\item {\bf Crowdsourcing and Research with Human Subjects}
    \item[] Question: For crowdsourcing experiments and research with human subjects, does the paper include the full text of instructions given to participants and screenshots, if applicable, as well as details about compensation (if any)? 
    \item[] Answer: \answerYes{} 
    \item[] Justification: Please see the Appendix F

\item {\bf Institutional Review Board (IRB) Approvals or Equivalent for Research with Human Subjects}
    \item[] Question: Does the paper describe potential risks incurred by study participants, whether such risks were disclosed to the subjects, and whether Institutional Review Board (IRB) approvals (or an equivalent approval/review based on the requirements of your country or institution) were obtained?
    \item[] Answer: \answerNA{} 
    \item[] Justification: -

\end{enumerate}

\end{document}